\documentclass[twocolumn]{aastex631}

\usepackage{graphicx} 
\usepackage{amsmath}
\usepackage[]{natbib}
\usepackage{color}

\graphicspath{{./}}

\begin{document}

\title{Bar instability and formation timescale across Toomre's $Q$ parameter and central mass concentration: slow bar formation or true stability}
\author{T. Worrakitpoonpon}
\affiliation{Institute of Science, Suranaree University of Technology, Nakhon Ratchasima 30000, Thailand}

\correspondingauthor{T. Worrakitpoonpon}
\email{worraki@gmail.com}
\date{}

\begin{abstract}
  We investigate the bar formation process using $N$-body simulations across the Toomre's parameter $Q_{min}$ and central mass concentration (CMC), focusing principally on the formation timescale. Of importance is that, as suggested by cosmological simulations, disk galaxies have limited time of $\sim 8$ Gyr in the Universe timeline to evolve secularly, starting when they became physically and kinematically steady to prompt the bar instability. By incorporating this time limit, bar-unstable disks are further sub-divided into those that establish a bar before and after that time, namely the normal and the slowly bar-forming disks.
  Simulations demonstrate that evolutions of bar strengths and configurations of the slowly bar-forming and the bar-stable cases are nearly indistinguishable prior to $8$ Gyr, albeit dynamically distinct, while differences can be noticed afterwards.
  Differentiating them before $8$ Gyr is possible by identifying the proto-bar, a signature of bar development visible in kinematical maps such as the Fourier spectrogram and the angular velocity field, which emerges in the former group $1-2$ Gyr before the fully developed bar, whereas it is absent in the latter group until $8$ Gyr and such bar-stable disk remains unbarred until at least $10$ Gyr.
  In addition, we find complicated interplays between $Q_{min}$ and CMC in regulating the bar formation. Firstly, disk stabilization requires both high $Q_{min}$ and CMC. Either high $Q_{min}$ or high CMC only results in slow bar formation. Secondly, some hot disks can form a bar more rapidly than the colder ones in a specific range of $Q_{min}$ and CMC.
\end{abstract}

\section{Introduction}
\label{sec:intro}

Barred galaxies constituted a branch of the Hubble diagram which accounted for more than $60\%$ of the observable disk galaxies in the present day \citep{eskridge_et_al_2000,marinova+jogee_2007,buta_et_al_2019}. That bar fraction was retained until the redshift $z\sim 1$ before it drastically decreased beyond that redshift \citep{jogee_et_al_2004,sheth_et_al_2008,melvin_et_al_2014}, dropping to $\sim 10\%$ at $z\sim 2$ \citep{guo_et_al_2023barjwst,le_conte_et_al_2024}. The search of the bar fraction at the earlier Universe remained a challenge, both theoretically and observationally, as only a couple of bars were discoverable at $z\sim 3$ \citep{constantin_et_al_2023}.
Not only in typical disk galaxies, bars were also found in low surface brightness galaxies \citep{peters_et_al_2019}, dwarf galaxies \citep{cuomo_et_al_2024}, and even in clustered galaxies  \citep{barazza_et_al_2009,lansbury_et_al_2014}.
Further examinations of the physical and kinematical properties of barred galaxies in details unveiled that bars could be found in galaxies of various mass \citep{nair+abraham_2010,erwin_2018}, bulge fraction 
\citep{vera_et_al_2016,barway_et_al_2016}, velocity dispersion \citep{sheth_et_al_2012,cervantes_sodi_2017}, and star formation rate \citep{lin_et_al_2017,newnham_et_al_2020}. This implied that the bar formation was a generic process across cosmic time and galaxy properties.

Bars were conjectured to stem from the global unstable two-armed modes in a locally stable disk according to the Toomre's criterion, designated by the Toomre's $Q$ parameter greater than $1$ \citep{toomre_1964,nipoti_et_al_2024}. 
That hypothesis has been tested and justified in pioneering simulations of isolated disks of various size and mass prone to the bar instability (see, e.g., \citealt{hohl_1971,miller+smith_1979,combes+sanders_1981}). 
Disk models were substantially elaborated in the years that followed by including the bulge or the halo coexisting with the disk in dynamical equilibrium so that the simulated systems were more relevant to the real galaxies, which greatly deepened the understanding in this field \citep{sellwood_1980}. 
For instance, it was reported that the concentration degree of the galactic bulge or the dark matter halo could be deterministic to the fate of the disk morphology such that a disk residing with concentrated components tended to form a bar slowly or be stable \citep{hohl_1976,norman_et_al_1996,rautiainen+salo_1999,sellwood+evans_2001,shen+sellwood_2004,athanassoula_et_al_2005,polyachenko_et_al_2016,kataria+das_2018,fujii_et_al_2018,saha+elmegreen_2018}. Likewise, high velocity dispersion in the disk could either slow down or stabilize the disk according to some studies \citep{jang+kim_2023,cas_paper}. In numerical aspect, it was documented that numerical artifacts such as the particle number, the integration time-step, and the softening length significantly affected the results, unless they were properly controlled \citep{dubinski_et_al_2009,frosst_et_al_2024}. The halo kinematical properties were also found important to the bar instability as it was concluded that the bar formation preferred the moderately spinning halo in the prograde direction to the prograde fast-rotating and the retrograde ones \citep{saha+naab_2013,long_et_al_2014,collier_et_al_2019b,lieb_et_al_2022, li_et_al_2023,joshi+widrow_2024}. On the other hand, the breaking of halo spherical symmetry was found to be another factor promoting the bar formation \citep{berentzen_et_al_2006,athanassoula_et_al_2013tri}. In the non-secular context, the global two-armed modes could otherwise originate from the external tidal disruption on a stable disk, namely the bridge-tail scenario \citep{toomre+toomre_1972,peschken+lokas_2019}.

While the understanding on the conditions that favored the bar formation has significantly been improved in the past decades, the importance of the bar formation timescale was not much regarded. The topic of the timescale can be equally important to the stability concern because the disk galaxies were supposedly allowed to evolve secularly in a limited time frame determined by the Universe history. 
It was marginally reported that the timescale of the secular bar formation depended on factors such as the disk fraction \citep{fujii_et_al_2018,bland_hawthorn_et_al_2023}, the bulge size \citep{kataria+das_2018}, the $Q$ parameter \citep{hozumi_2022}, and the thickness \citep{ghosh_et_al_2023,ghosh_et_al_2024}. In those studies, the formation timescales ranged from $1$ Gyr to as long as $10$ Gyr. 
In the cosmological framework, various suites of $\Lambda$CDM cosmological simulations suggested that the bar formation was not monolithic as it could be initiated at any time from the redshift $2$ to the recent days \citep{zhou_et_al_2020,izquierdo_villalba_et_al_2022}. 
These facts suggest that there should be a number of galaxies that appear unbarred currently but the bar formation is ongoing, due to either the recent onset or the slow progress of the bar formation. 
As a matter of fact, the examination of the conditions that give rise to those disk galaxies and their observable properties that can be distinguished from the stable ones can be pivotal to the current understanding.

In this work, we analyze the bar instability in the enlarged scope compared to past studies. We not only examine the bar instability across the range of the disk and the halo parameters, but we also explore the bar formation timescales and specify the disks that form a bar slowly, i.e., those that cannot become fully barred before the time permitted by the Universe timeline. These disks can be misidentified as stable due to slow bar formation if they are not evolved long enough, and we speculate that they constitute a significant fraction among observable unbarred galaxies. 
The article is organized as follows. First of all, Sec. \ref{sec:nume} describes the galaxy model employed in this study, the details of the numerical simulations, and the bar parameters. Then, Sec. \ref{sec:bar_1} presents the analyses of the bar formation and their timescales for different cases and their dependences on the system parameters. We mainly focus on cases that form a bar slowly and the capability to distinguish them from the stable ones. In Sec. \ref{sec:bar_crit} that follows, we revisit the stability criterion in the spaces of the system parameters, involving the fact that there exists slowly bar-forming disks. Finally, Sec. \ref{sec:concls} is for the conclusion.

\section{Simulation details and parameters}
\label{sec:nume}

\subsection{Galaxy model}
\label{ssec:gal_mod}
The density of the disk of particles follows the exponential profile with finite thickness as follows
\begin{equation}
\rho_{d}(r,z)=\frac{M_{d}}{4\pi R_{0}^{2}z_{0}}e^{-r/R_{0}}\text{sech}^{2}{\bigg(\frac{z}{z_{0}}\bigg)}
    \label{density_disk}
\end{equation}
where $M_{d}$ is the disk mass, $R_{0}$ is the disk scale radius, and $z_0$ is the disk scale thickness. We choose $M_{d}=10^{10} \ M_{\odot}$, $R_{0}=5 \ \text{kpc}$, and $z_{0}=0.2 \ \text{kpc}$. The disk is radially and vertically truncated at $5R_{0}$ and $5z_{0}$, respectively. Disks are put in a spherically symmetric dark matter halo of particles (or a live halo) having the Hernquist density profile \citep{hernquist_1990} given by
\begin{equation}
\rho_{h}(r)=-\frac{M_{h}r_{h}}{2\pi r(r+r_{h})^{3}},
    \label{den_hern}
\end{equation}
where $M_{h}$ and $r_{h}$ are the halo mass and the halo scale radius. We fix the halo mass to $25M_{d}$ while we vary $r_{h}$ for different values of the central mass concentration (hereafter CMC). The CMC (or $\mathcal{C}$) corresponds to the total mass enclosed inside $0.2 \ \text{kpc}$ relative to the disk mass. For convenience, the CMC values multiplied by $10^{3}$ are those for analysis. 
The halo is truncated at $2r_{h}$. The radial $Q$ profile corresponds to the ratio of the radial velocity dispersion $\sigma_{r}$ to the critical value for the local stability following the Toomre's criterion \citep{toomre_1964}, i.e.,
\begin{equation}
    Q=\frac{\sigma_{r}\kappa}{3.36G\Sigma}
    \label{q_def}
\end{equation}
where $G$ is the gravitational constant; $\kappa$ is the epicyclic frequency calculated from the composite disk-halo potential $\Phi_{tot}$  by
\begin{equation}
\kappa^{2}=\frac{d^{2}\Phi_{tot}}{dr^{2}}+\frac{3}{r}\frac{d\Phi_{tot}}{dr}; 
\label{eq:kappa}
\end{equation}
and $\Sigma$ is the disk surface density derived from the density profile (\ref{density_disk}), which is proportional to $e^{-r/R_{0}}$. We adopt the prescriptions of \citet{hernquist_1993} for the velocity structures of the disk-halo system in the dynamical equilibrium, which can be detailed as follows. Based on observations, the radial velocity dispersion relates to the surface density $\Sigma$ as
\begin{equation}
\sigma_{r}^{2}\propto \Sigma. \label{sigmar_prop}
\end{equation}
We choose the minimum $Q$, namely $Q_{min}$, to represent the initial kinematical condition of a disk. The tangential velocity dispersion as a function of radius $\sigma_{\theta}$ is given by
\begin{equation}
\sigma_{\theta}^{2}= \frac{\kappa^{2}}{4\Omega^{2}}\sigma_{r}^{2}, \label{sigmatheta}
\end{equation}
where $\Omega$ is the angular frequency of circular orbit computed from $\Phi_{tot}$ as
\begin{equation}
\Omega^{2}=\frac{1}{r}\frac{d\Phi_{tot}}{dr}.
\label{eq:omega}
\end{equation}
The vertical velocity dispersion $\sigma_{z}$ can be obtained from 
\begin{equation}
\sigma_{z}^{2}= \pi G\Sigma z_{0}. \label{sigmaz}
\end{equation}
The mean tangential velocity $\bar{v}_{\theta}$ is deduced from the axisymmetric Jeans equation, namely
\begin{equation}
\bar{v}_{\theta}^{2}=r^{2}\Omega^{2}+\frac{r}{\Sigma}\frac{d(\sigma_{r}^{2}\Sigma)}{dr}+\sigma_{r}^{2}-\sigma_{\theta}^{2},
\label{vthetamean}
\end{equation}
whereas the mean radial and vertical velocities (or $\bar{v}_{r}$ and $\bar{v}_{z}$) are set to zero. For the velocity structure of the live halo, the first moments of the velocities are zero. We assume the velocity isotropy so that the velocity dispersions of all axes are identical, and the magnitude as a function of radius $\sigma_{h}$ is derivable from the spherically symmetric Jeans equation
\begin{equation}
    \sigma_{h}^{2}=\frac{1}{\rho_{h}}\int_{r}^{\infty}\frac{GM_{tot}\rho_{h}}{r^{2}}dr,
\label{sigma_halo}    
\end{equation}
where $M_{tot}$ is the total mass enclosed inside $r$. Disk and halo random velocity components are drawn from the cut-off Gaussian distribution. 

\begin{figure}
    \centering
    \includegraphics[width=9.0cm]{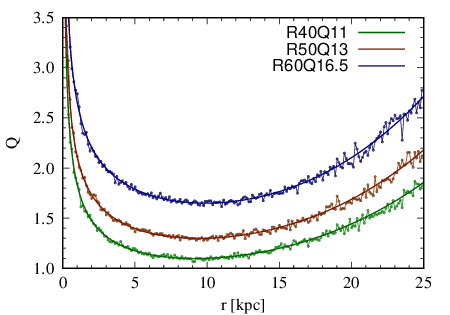}
    \caption{Initial radial $Q$ profiles for different initial states. Solid lines represent theoretical values whereas connected points are the measured values from initial $N$-body disks.}
    \label{fig:profil_q}
\end{figure}

In all simulations, the disk and the halo consist of $2\times 10^{6}$ and $3\times 10^{6}$ particles, respectively. The details of the parameters for different initial conditions and their code names are given in Tab. \ref{tab_ini}. The case name consists of two parts designating the halo scale radius and $Q_{min}$. The numerical value after R represents $r_{h}$ in kpc, whereas the value after Q stands for $10Q_{min}$.
We examine 4 different levels of the CMC from 4 halo sizes, which can be categorized into R40, R50, R60 and R75. Otherwise, when classified by $Q_{min}$, we have Q11, Q12, Q13, Q14, Q15, and Q16.5 families. The value of $Q_{min}=1.65$ is almost the upper limit for physical solution of the Jeans equation (\ref{vthetamean}) and the solutions are available only for $r_{h}\lesssim 60$ kpc. Our choice of $\sigma_{r}$ yields the characteristic U-shape radial $Q$ profile, illustrated in Fig. \ref{fig:profil_q} for real and measured values for different cases, which is analogous to other studies. Different $Q_{min}$ can be obtained by varying the pre-factor in Eq. (\ref{sigmar_prop}), and the profile shifts accordingly with $Q_{min}$. We note that $Q_{min}$ is situated close to $2R_{0}$ and the value of $Q\sim Q_{min}$ spans a considerable range in the disk interior. More specifically, the $Q$ value exactly at $2R_{0}$ differs from $Q_{min}$ by $0.3 \ \%$ at most.

\begin{table}
    \caption{Parameters of initial disks and their code names}
    \centering
    \begin{tabular}{l|c|c|c}
        \hline \hline   
        Code name & $r_{h}$ (kpc) & $\mathcal{C}$ ($\times 10^{-3})$ & $Q_{min}$ \\
        \hline \hline   
        R40Q11 & 40 & 1.398 & 1.1 \\
        R40Q12 & 40 & 1.398 & 1.2 \\
        R40Q13 & 40 & 1.398 & 1.3 \\
        R40Q14 & 40 & 1.398 & 1.4 \\
        R40Q15 & 40 & 1.398 & 1.5 \\
        R40Q16.5 & 40 & 1.398 & 1.65 \\
        \hline
        R50Q11 & 50 & 1.176 & 1.1 \\
        R50Q12 & 50 & 1.176 & 1.2 \\
        R50Q13 & 50 & 1.176 & 1.3 \\
        R50Q14 & 50 & 1.176 & 1.4 \\
        R50Q15 & 50 & 1.176 & 1.5 \\
        R50Q16.5 & 50 & 1.176 & 1.65 \\
        \hline
        R60Q11 & 60 & 1.055 & 1.1 \\
        R60Q12 & 60 & 1.055 & 1.2 \\
        R60Q13 & 60 & 1.055 & 1.3 \\
        R60Q14 & 60 & 1.055 & 1.4 \\
        R60Q15 & 60 & 1.055 & 1.5 \\
        R60Q16.5 & 60 & 1.055 & 1.65 \\
        \hline
        R75Q11 & 75 & 0.956 & 1.1 \\
        R75Q12 & 75 & 0.956 & 1.2 \\
        R75Q13 & 75 & 0.956 & 1.3 \\
        R75Q14 & 75 & 0.956 & 1.4 \\
        R75Q15 & 75 & 0.956 & 1.5 \\
        \hline
    \end{tabular}
    \label{tab_ini}
\end{table}

\subsection{Simulation of dynamics of galaxies}
\label{ssec:sim_detail}
The equations of motions of particles are integrated using GADGET-2 \citep{springel_2005}. The gravitational force is spline-softened by a predefined softening length which is fixed to $5$ pc for all particles. Calculations of mutual forces between particles are facilitated by the tree code with an opening angle fixed to $0.7$ which is also applied to all particles. The integration time step is controlled to be not greater than $0.1 \ \text{Myr}$, yielding the accuracy such that the deviations of the total energy and the total angular momentum at the end of simulation are not greater than $0.15 \%$ and $0.2 \%$ of the initial values, respectively. 

\subsection{Bar mode parameters}
\label{ssec:bar_param}

Eminences of bi-symmetric features in a disk can be evaluated by the Fourier amplitude as a function of radius $\tilde{A}_{2}(r)$ defined as
\begin{equation}
    \tilde{A}_{2}(r)=\frac{\sqrt{a^{2}_{2}+b^{2}_{2}}}{A_{0}}
    \label{tilde_a2}
\end{equation}
where $a_{2}$ and $b_{2}$ are the $m=2$ mode Fourier coefficients at $r$, and $A_{0}$ is the corresponding $m=0$ mode amplitudes. In practice, it is calculated from the particles inside the annulus of radius $r$. The bar strength $A_{2}$ is designated by the maximum $\tilde{A}_{2}$ within $r_{max}$, namely
\begin{equation}
    A_{2}\equiv \max_{r<r_{max}}[\tilde{A}_{2}].
    \label{a2_def}
\end{equation}
The conventional threshold value of $A_{2}=0.2$ is employed to classify between barred and unbarred states for our investigation and a disk is considered fully barred when $A_{2}$ reaches and remains well above $0.2$. The bar angular alignment can be obtained from the phase of the Fourier amplitudes calculated from all particles inside $r_{max}$, namely $a_{2,tot}$ and $b_{2,tot}$, as
\begin{equation}
    \phi_{2}\equiv \frac{1}{2}\tan^{-1}\bigg( \frac{b_{2,tot}}{a_{2,tot}}\bigg).
    \label{phi2_glob_def}
\end{equation}
In the formulae (\ref{a2_def}) and (\ref{phi2_glob_def}), we fix $r_{max}$ to $10$ kpc. The reason for which we choose $r_{max}$ slightly beyond the bar extent is to exclude the spiral pattern from calculations.

\section{Timescales of bar formation and disk kinematics}
\label{sec:bar_1}

\subsection{Bar instability, slow bar formation or true stability}
\label{ssec:bar_late}
\begin{figure*}
    \centering
    \includegraphics[width=18.0cm]{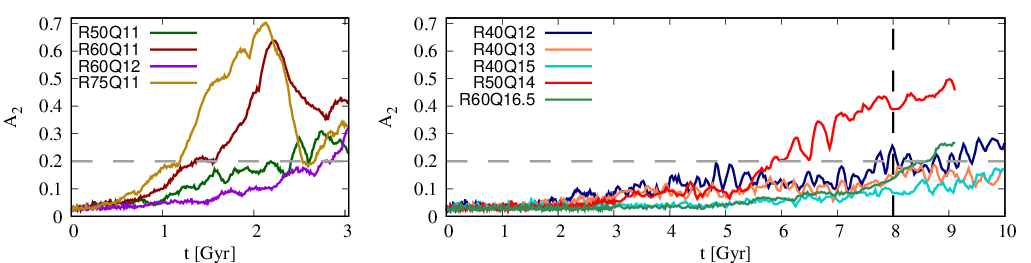}
    \caption{Time evolutions of $A_{2}$ for different indicated cases. The horizontal dashed line designates the value of $A_{2}=0.2$, serving as the threshold for the fully barred state. The vertical dashed line indicates the time at $8$ Gyr, used for the distinction between normal and slow bar formations.}
    \label{fig:a2}
\end{figure*}

\begin{figure*}
    \centering
    \includegraphics[width=18.0cm]{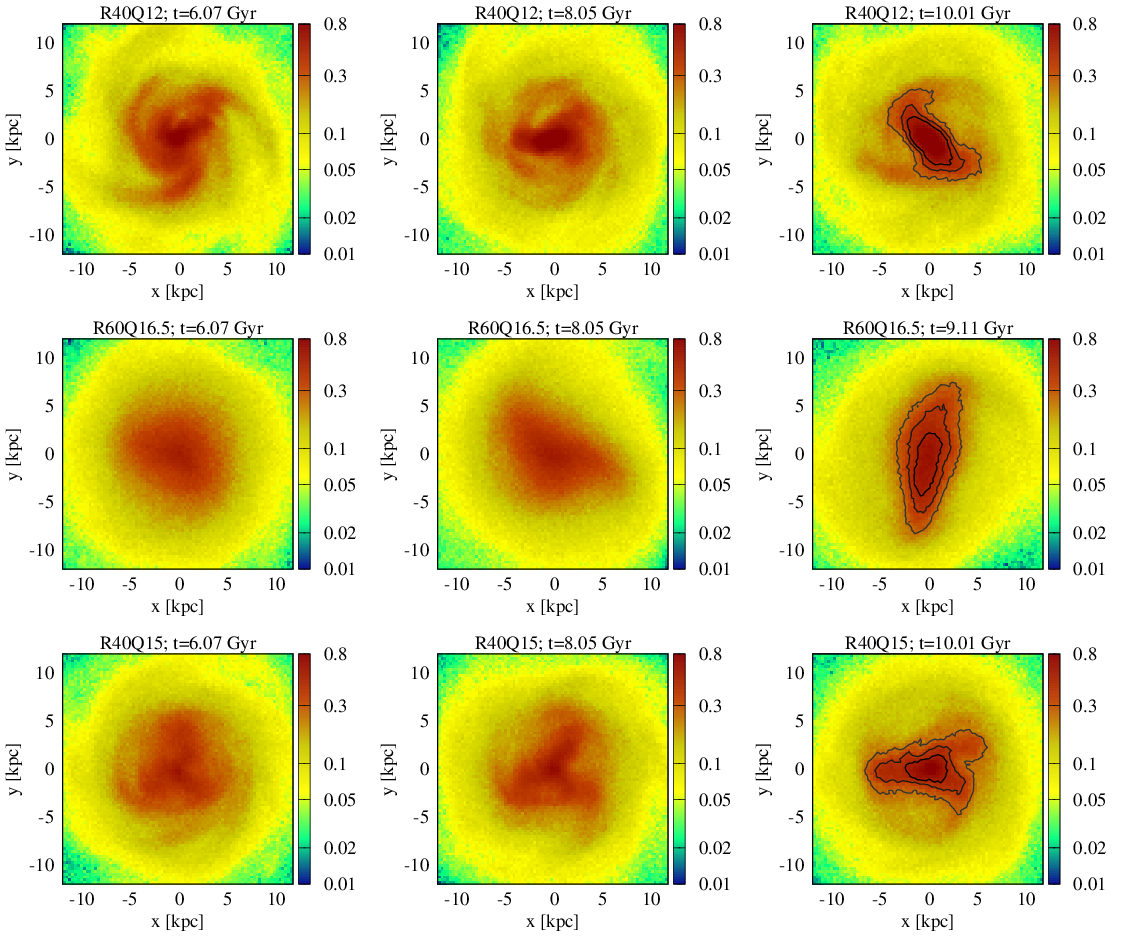}
    \caption{Disk surface densities in color map for R40Q12 (top row), R40Q15 (middle row) and R60Q16.5 (bottom row) at indicated times. Surface densities are in units of $10^{8} \ M_{\odot}/$kpc$^{2}$. The isodensity contours in the last snapshot for each case represent, from outside to inside, the values of $0.4\Sigma_{0}, 0.6\Sigma_{0}$ and $0.8\Sigma_{0}$, where $\Sigma_{0}\equiv M_{d}/2\pi R_{0}^{2}$ corresponding to the central surface density of the initial disk.}
    \label{fig:snap_late}
\end{figure*}

We first of all examine the evolution of $A_{2}$ for disks of various $Q_{min}$ and $\mathcal{C}$, as shown in Fig. \ref{fig:a2} for some selected cases. Comparing the bar formation timescales between cases, it is evident that increasing $Q_{min}$ or $\mathcal{C}$ tends to slow down the bar formation. The 4 cases in the left panel which are among the cases with low $Q_{min}$ and CMC, form a bar within $3$ Gyr. The R50Q14 case which has elevated $Q_{min}$ and $\mathcal{C}$, becomes barred around $6$ Gyr. Slower bar formations are spotted for R40Q12 and R60Q16.5 as $A_{2}$ surpasses and remains above $0.2$ after $8$ Gyr. To verify if the bars are really established, the configurations of the two slowly bar-forming cases are depicted in Fig. \ref{fig:snap_late}. 
The R60Q16.5 case at $6.07$ Gyr does not yet exhibit a bar as only a mildly deformed disk center is observed, in accordance with the value of $A_{2}$ that merely attains $0.1$. 
At the end, the assembly of the bar of $\sim 8$ kpc of length is verified, as evaluated by the extent of the $0.4\Sigma_{0}$ isodensity contour (see, e.g., \citealt{pfenniger+friedli_1991,michel-dansac+wozniak_2006} for examples of detailed analysis of bar anatomies using the isodensity contours). The R40Q12 case exhibits a shorter bar with surrounding ring-like structure in addition which is the remnant of the shearing by a high CMC.
On the other hand, R40Q13 and R40Q15 prove stable against bi-symmetric perturbations until $10$ Gyr as $A_{2}$ slightly increases from the initial value and saturates at $\sim 0.15$ until the end.
The stability against bar modes for R40Q15 is justified by the $0.4\Sigma_{0}$ isodensity line (see Fig. \ref{fig:snap_late}) which has the multi-arm appearance until $10$ Gyr. The formation of the spiral structure is attributed to the mechanism of the swing amplification, an amplification of non-axisymmetric spiral modes by the synchronized epicyclic motion in combination with the disk shearing \citep{julian+toomre_1966,de_rijcke_et_al_2019,michikoshi+kokubo_2020}.
The shearing degree of the R40 family proves capable of stabilizing some disks with intermediate to high $Q_{min}$, by means of neutralizing the bar modes and yielding the persistent spiral structure in place. Contrarily, the highest $Q_{min}=1.65$ cannot subdue the bar instability if $\mathcal{C}$ is low as it only delays the process to the point that the bar is fully developed after $8$ Gyr. The stability criterion in parameter spaces and the interplay between different mechanisms in regulating the bar formation will be examined more in Sec. \ref{sec:bar_crit}.

In the morpho-dynamical perspective, all cases presented in Fig. \ref{fig:a2} except R40Q13 and R40Q15 can be classified as bar-unstable. When we impose the time limit in which disks are permitted to evolve, some of them can be disguised as bar-stable. For instance, if R40Q12, R40Q13, R40Q15 and R60Q16.5 are only evolved to $8$ Gyr or earlier, some of them can be mistakenly classified as bar-stable because the evolutions of $A_{2}$ for the four cases are nearly indistinguishable prior to that time limit. The difference in the evolution pattern can be observed afterwards. If we include the possibility of the slow bar formation, the scope of the bar instability should be revised.

Before we proceed on this subject, we firstly ensure that our timescales are compatible with other studies of isolated disk-halo systems (see, for instance, \citealt{kataria+das_2018,fujii_et_al_2018,kataria+das_2019,bauer+widrow_2019,jang+kim_2023}) and disks in cosmological simulations \citep{lokas_2021bar,marioni_et_al_2022}.
Secondly, the time limit of $8$ Gyr to distinguish between the normal and the slow bar formations, which has been included in Fig. \ref{fig:a2}, needs supporting reasons.
Studies of barred-galaxy population in suites of cosmological simulations (see, e.g., \citealt{zhou_et_al_2020,izquierdo_villalba_et_al_2022}) reported the median times of the emergences of the bars in the range from $4-7$ Gyr of the lookback time. Including the time period of $1-3$ Gyr from the onset, as suggested by our results for typical low-$Q_{min}$ disks, justifies that the onset of the bar growth could averagely be marked at $8$ Gyr ago, in agreement with other analyses that tracked the secular bar growth individually \citep{lokas_2021bar,marioni_et_al_2022}. 
In the observational context, estimates of the onset time could be performed  using the age of the nuclear disk \citep{baba+kawata_2020,de_sa_freitas_et_al_2023,sanders_et_al_2024} or the age of a specific star type \citep{cole+weinberg_2002,james+percival_2016}, concluding that the onsets were initiated $3-10$ Gyr ago.
Other supports were from the analyses of galaxy properties in large-scale cosmological simulations which suggested that typical disk galaxies were steady, as probed by the mass \citep{lokas_2021bar}, the angular momentum \citep{grand_et_al_2017}, the rotational-to-random velocity ratio \citep{jackson_et_al_2020a}, and the bulge-to-total ratio \citep{zeng_et_al_2021,zana_et_al_2022}, not before $8$ Gyr ago so that the bar growth could be initiated. 
Bar growths prior to that time were attributed to the galaxy encounter or merger, implanting an asymmetric seed so that the bar growth was able to be triggered by $\sim 2$ Gyr in advance \citep{algorry_et_al_2017,rosas_guevara_et_al_2020,lopez_et_al_2024}. Those barred galaxies typically started with bar amplitudes already close to $0.2$.

Including the time limit into the morpho-dynamical perspective, we can define a sub-category for bar-unstable disk galaxies which cannot establish a bar before $8$ Gyr but they manage to do so within a couple of Gyr, namely the slowly bar-forming category. These disks exhibit slow continual growth of the bar amplitude and finally reach the barred state after $8$ Gyr, which clearly differ from the bar-stable counterparts whose $A_{2}$ saturates well below $0.2$ until $10$ Gyr, owing to the strong shearing. On the other hand, disks that are able to form a bar before $8$ Gyr are simply classified as normal bar-unstable cases.
This finding can be important to the field of extragalactic astrophysics as we speculate that the slowly bar-forming disks constitute a fraction of currently unbarred galaxies, which will become barred in a few Gyr. Dynamics-wise, they should not be classified into the same category as the bar-stable disks. The method to identify the slowly bar-forming galaxies will be addressed in Sec. \ref{ssec:kine_forming}.

\subsection{Physical and kinematical structures indicative to slow bar formation}
\label{ssec:kine_forming}

\begin{figure*}
    \centering
    \includegraphics[width=18.0cm]{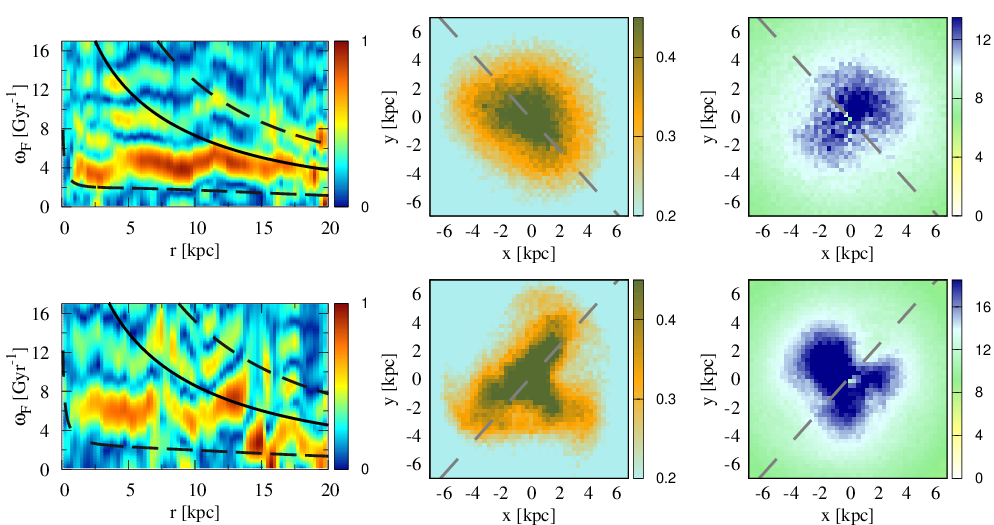}
    \caption{Fourier spectrograms as a function of radius and frequency $\omega_{F}$ (left panels), zoom-in surface densities in a fine color scale (middle panels), and angular frequency fields (right panel). All figures in the top row are for R60Q16.5, whereas those in the bottom row are for R40Q15. The Fourier spectrograms are calculated in the time window from $4.86-6.07$ and $7.59-8.50$ Gyr for R60Q16.5 and R40Q15, respectively. Fourier spectra are presented in units of the maximum value in each plot. The solid line represents the angular frequency $\Omega$, while the dashed lines above and below $\Omega$ correspond to $\Omega+\kappa/2$ and $\Omega-\kappa/2$, respectively, where $\kappa$ is the epicyclic frequency. All frequencies are calculated from the total disk-halo potential. For the fine-scale surface densities and the angular frequency fields, they are taken at $6.07$ and $8.05$ Gyr for R60Q16.5 and R40Q15. The dashed lines represent the bar phases at those times. The units of the density are the same as in Fig. \ref{fig:snap_late}, while the units of the angular frequencies are $\text{Gyr}^{-1}$.}
    \label{fig:fou_spec}
\end{figure*}

As demonstrated in Sec. \ref{ssec:bar_late}, the evolutions of $A_{2}$ for the cases that do not form a bar and those that form a bar slowly could not be distinguished properly prior to $8$ Gyr which is the time limit for the secular evolution of a disk according to recent cosmological simulations. Moreover, if the fates of R40Q15 and R60Q16.5, for instance, are not perceived beforehand, the configurations in the midst of evolution, namely at $6.07$ Gyr, are not well informative to whether the disk is stable or slowly bar-forming as both only exhibit mildly deformed disk centers from circular symmetry (see Fig. \ref{fig:snap_late}).
Therefore, finding a reliable way to differentiate between these two disks prior to $8$ Gyr is an important milestone towards better understanding of the bar formation mechanism.
In this section, we inspect further the kinematical details of the disk interiors, specifically those that are slowly bar-forming and stable, namely R60Q16.5 and R40Q15. We frame the time at which $A_{2}\sim 0.1$ for both cases in our investigation. To do so, we plot the Fourier spectrograms of the $m=2$ modes as a function of radius and frequency $\omega_{F}$, the zoom-in surface density maps in a finer color scale, and the angular frequency fields for both cases in Fig. \ref{fig:fou_spec}. The Fourier spectrograms are computed during $4.86-6.07$ and $7.59-8.50$ Gyr for R60Q16.5 and R40Q15, respectively. We include the curves for $\Omega$ (solid lines) and $\Omega\pm\kappa /2$ (dashed lines), all of which are calculated from the total disk-halo potential. For the surface densities, we re-plot them in a finer color scale to discriminate subtle variation of the density more appropriately. The angular frequency fields are for inspecting the particle angular motion in response to the growing non-axisymmetric modes. The two latter plots are taken at $6.07$ Gyr for R60Q16.5; and $8.05$ Gyr for R40Q15. The dashed lines in the two latter plots represent the bar phases at those times.

The Fourier spectrograms for the two disks display completely different features as for R60Q16.5, we observe the dominant $m=2$ component with almost uniform frequency of $\sim 5 \ \text{Gyr}^{-1}$ that continuously spans $\sim 15$ kpc and intersects with the $\Omega$ line. This implies that the corotation resonance is responsible for the development of the bar. Such structure emerges at $\sim 2$ Gyr before the fully developed bar and remains until that time. The presence of the bi-symmetric rigidly-rotating structure in the Fourier spectrogram is in coherence with the mild bar-like appearance observed in the refined surface density plot, which accordingly aligns with the bar phase. We designate this component as the 'proto-bar' because it is not yet eminent in the disk compared to the fully developed bar (see Fig. \ref{fig:snap_late}), but it manifests sign of the bar growth, able to withstand the shearing until it is fully developed.
The presence of the proto-bar leads to the characteristic persistent two-lobed feature in the angular frequency map that aligns perpendicularly to the bar axis. This can be interpreted that particles along the bar axis have lower angular frequencies closer to the $m=2$ pattern speed than the off-axis ones, which signifies the ongoing particle trapping by the bar potential. Among the slowly bar-forming cases, the lifetime of the proto-bar is typically around $1$ Gyr, and it is $2$ Gyr at most as for the representative case.
The proto-bar is, on the contrary, absent in the plots for R40Q15. As displayed in the Fourier spectrogram, we rather observe scattered components of various angular frequencies, with a fragment overlaid with the $\Omega$ line. This means that there is no bar development by the corotation resonance as with the R60Q16.5 counterpart. The scattered components could be the result of the shearing that tears the bar modes into pieces of various angular frequencies. The refined surface density plot which is measured at $\sim 8$ Gyr, has the feature irrelevant to the proto-bar. Instead, it exhibits the long-lasting multi-arm structure in line with the angular frequency map. 

\begin{figure*}
    \centering
    \includegraphics[width=18.0cm]{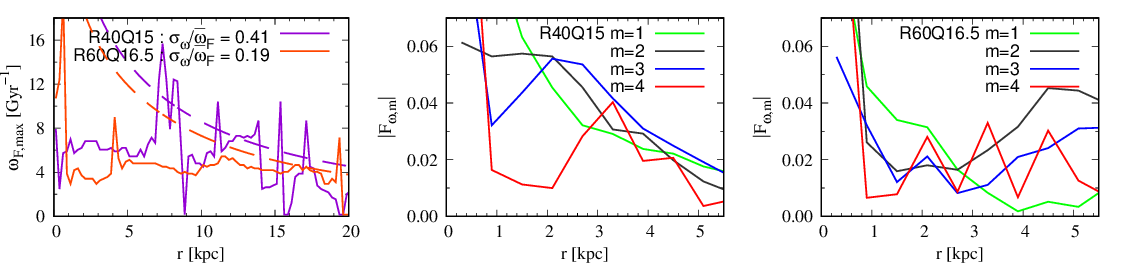}
    \caption{Dominant Fourier frequency $\omega_{F,max}$ as a function of radius for R40Q15 and R60Q16.5 in solid lines, with the disk angular frequencies $\Omega$ in dashed lines with the same colors (left panel); and Fourier $m$-mode amplitude as a function of radius for the angular velocity field for R40Q15 (left panel) and R60Q16.5 (right panel). Plots for each case are taken at the times in Fig. \ref{fig:fou_spec}.}
    \label{fig:profil_kinetic}
\end{figure*}

In continuity with the two Fourier spectrograms, we extract the dominant frequency $\omega_{F,max}$ as a function of radius and it is plotted in the left panel of Fig. \ref{fig:profil_kinetic} for the two cases, with the disk angular frequency $\Omega$ alongside for the consideration of the resonance. As implied by the Fourier spectrograms, $\omega_{F,max}$ for R60Q16.5 is more uniform than the R40Q15 counterpart. In order to discriminate between the two cases, we propose the frequency dispersion to the mean frequency ratio, namely $\sigma_{F}/\bar{\omega}_{F}$, measured in the disk interior to the radius susceptible to the resonance point. As a result, the ratios for R40Q15 and R60Q16.5, measured from $2.5-15$ kpc, are equal to $0.41$ and $0.19$, respectively. It turns out that the value $\sigma_{F}/\bar{\omega}_{F}=0.3$ can distinguish between the proto-bar and the multi-arm pattern. 
Otherwise, we repeat the radial Fourier $m$-mode analysis as for the bar amplitude $\tilde{A}_{2}(r)$ but now we apply to the angular frequency field in Fig. \ref{fig:fou_spec} to obtain the corresponding radial Fourier $m$-mode amplitude profile, namely
\begin{equation}
    |F_{\omega,m}(r)|=\Bigg| \frac{\int_{0}^{2\pi}\varpi (r,\theta) e^{im\theta}d\theta}{\int_{0}^{2\pi}\varpi (r,\theta) d\theta} \Bigg|,
\end{equation}
where $\varpi (r,\theta)$ is the angular frequency at $(r,\theta)$. Different modes are then depicted in the middle and the right panels for R40Q15 and R60Q16.5, respectively. Although all modes are still weak, it is evident that the $m=2$ modes dominate the other modes for R60Q16.5, whereas for R40Q15, the $m=2$ modes are topped by the $m=3$ modes and they are comparable to the $m=1$ counterparts. The Fourier analysis of the angular frequency field is another tool for identifying the proto-bar beside the radial profile of $\omega_{F,max}$. In connection with the observation, both profiles can be obtained by converting the rectilinear velocity field which can be decomposed into the radial and tangential components by proper techniques \citep{wu_et_al_2021,wolfer_et_al_2023,sylos_labini_et_al_2023}, using the radial distance from the galaxy center. On the other hand, to inspect the coherence of the pattern speed in search of the proto-bar, the angular frequency can be narrowed down to be along an axis susceptible to hosting a proto-bar, as performed by \citet{treuthardt_et_al_2007,aguerri_et_al_2015,geron_et_al_2023}. 
The analysis on the kinematical maps is complementary to the conventional morphological analysis on the surface brightness as it is based on the spectroscopic data. Ideally, the genuine proto-bar should exhibit features in both the photometric and the kinematical maps, but we are inclined to the kinematical analysis as more favorable in identifying the proto-bar since it reflects the actual dynamical picture of the disk more appropriately.
In continuity with the proposed classification incorporating the time limit in Sec. \ref{ssec:bar_late}, the presence of the proto-bar prior to $8$ Gyr is an important factor differentiating the yet unbarred galaxies. More precisely, the slowly bar-forming disks are those that host the proto-bar before that time and they are likely to form a bar within $8-10$ Gyr, whereas the bar-stable ones do not. Our proposed classification scheme for unbarred galaxies is enlarged from past schemes which were based on the configuration parameters, the photometric properties, or the morphologies only. We demonstrate that with some additional kinematical maps, the unbarred galaxies can be further classified as either the slowly bar-forming or the stable galaxies.

\begin{figure*}
    \centering
    \includegraphics[width=18.0cm]{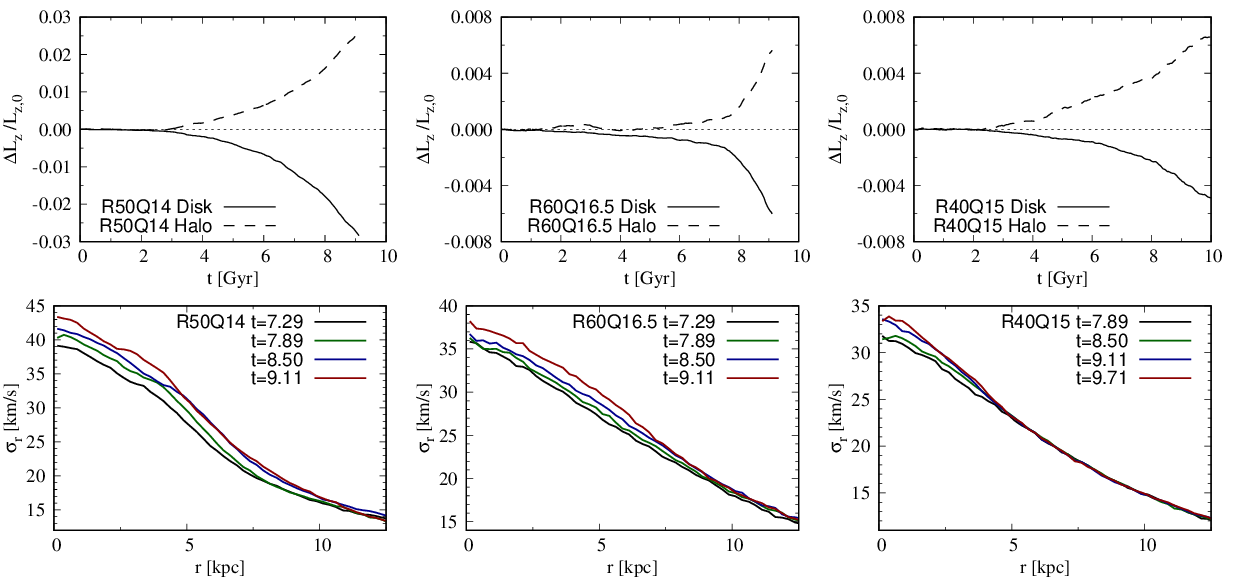}
    \caption{Change of $z$-axis angular momentum for disk (solid lines) and halo (dashed line) relative to the initial total angular momentum $\Delta L_{z}/L_{z,0}$ (top panels) and radial velocity dispersion profile at different times (bottom panels) for R50Q14 (left panels), R40Q15 (middle panels) and R60Q16.5 (right panels).}
    \label{fig:dispr_moment}
\end{figure*}

Bar formation and evolution have always been associated with the disk-halo angular momentum transfer and the radial heating. To verify if such processes can be employed to distinguish between the normal bar-forming, the slowly bar-forming and the bar-stable ones, time evolutions of the disk and the halo angular momentum changes relative to the initial total ones, namely $\Delta L_{z}/L_{z,0}$, and the radial velocity dispersion profiles $\sigma_{r}$ at different times for R50Q14, R40Q15 and R60Q16.5 are plotted in Fig. \ref{fig:dispr_moment}. Considering firstly the angular momentum plot, $\Delta L_{z}/L_{z,0}$ for R50Q14 surpasses $2\%$ at the end, which is more than $3$ times of those for R60Q16.5 and R40Q15, although the formation timescales between R50Q14 and R60Q16.5 only differ by $2$ Gyr. We conclude that the normal bar formation leads to significantly higher amount of $\Delta L_{z}/L_{z,0}$ than the two other groups. On the contrary, the differentiation between the two other cases is difficult as they manifest comparably weak angular momentum exchanges.
This can be explained that the angular momentum transfer requires the resonances between non-axisymmetric modes and disk natural frequencies which are not limited to the bar modes \citep{athanassoula_2003,holley-bockelmann_et_al_2005,villa-vargas_et_al_2009,minchev_et_al_2012,athanassoula_et_al_2013tri,saha+jog_2014,chiba+schonrich_2022}. Therefore, weak spiral modes can lead to weak angular momentum transfer as with the proto-bar.
On the other hand, the process of the radial heating proves to be an inefficient classifier as the evolutions of $\sigma_{r}(r)$ for the three cases do not differ significantly. The central $\sigma_{r}$ for R50Q14 is slightly more heated up than the two other cases in the course of the final $2$ Gyr. The heating is more widespread for R50Q14 as the profile up to $10$ kpc systematically shifts up, compared to the two other cases where the heating takes place well inside $10$ kpc.
These can be explained by the similar arguments: the radial heating is not limited to the bar modes but other asymmetric modes such as the spiral arms \citep{jenkins+binney_1990,thomasson_et_al_1991,minchev+quillen_2006,minchev_et_al_2012,gustafsson_et_al_2016,seo_et_al_2019}, or even the clumps \citep{sanchez-salcedo_1999,hanninen+flynn_2002,ardi_et_al_2003,vande_putte_et_al_2009,fujimoto_et_al_2023,cas_paper}, can also be responsible for it.

The term 'proto-bar' has indeed been coined in the work of \citet{zana_et_al_2022} who analyzed the barred galaxy population in the cosmological simulation. Their statistical surveys specified the proto-bar based on the value of $A_{2}$ that fell between $0.1-0.2$. According to our results, that does not suffice as the stable case that harbors spiral arms, can produce the value of $A_{2}$ comparable to that for the real proto-bar and such level of $A_{2}$ can be maintained for many Gyr. The differentiation between the spiral structure and the genuine proto-bar in the stable and the slowly bar-forming disks is possible by the kinematical analyses in the Fourier spectrogram and the angular frequency field in addition to the fine-scale density map, covering the period susceptible to hosting the proto-bar. 

The concept of the proto-bar can be exploited in the observational context in the following way. If the proto-bar is spotted in a currently unbarred galaxy, it is likely to become a barred galaxy in a few Gyr. Note that the proto-bar is not only specific to the slowly bar-forming disks, but disk galaxies that form a bar rapidly also have to undergo the proto-bar stage. However, to pinpoint the proto-bar for cases with fast bar formation might not be that practical and useful as, firstly, the lifetime of the proto-bar stage might be too short. Our representative case with slow bar formation exhibited the longest proto-bar lifetime which is $\sim 2$ Gyr prior to the time of the fully developed bar. This suggests that the proto-bar lifetime should be approximately one-forth of the formation timescale at most. Therefore, cases that form a bar rapidly only spend a fraction of Gyr in the proto-bar stage. Secondly, disks that maintained the stability against bar modes until recently were not likely to trigger the fast bar formation secularly. Rather, bar formation should be from external cause such as the tidal interaction which developed a bar rapidly beyond the pertubative formalism, making the presence of the proto-bar stage questionable. In summary, the inspection of the proto-bar is possible if the bar formation is not too rapid and the period of the proto-bar stage coincides with the time of the observation. However, the capture of the proto-bar might be limited to nearby galaxies as it requires sufficient resolution of the disk plane for fine details of the density contrast and the local kinematics.

\subsection{Bar growth rate as a function of $Q_{min}$ and CMC}
\label{ssec:growth_rate}

\begin{figure}
    \centering
    \includegraphics[width=9.0cm]{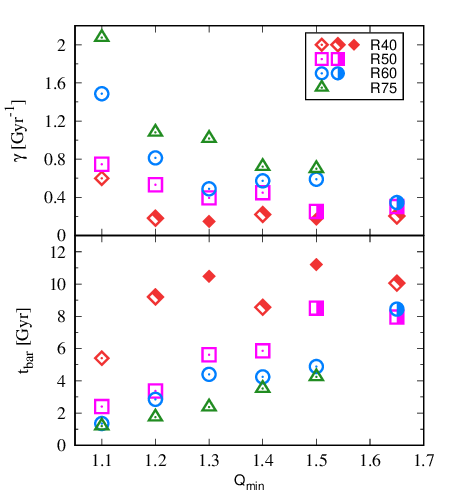}
    \caption{Bar growth rate $\gamma$ (top panel) and formation timescale $t_{bar}$ (bottom panel) as a function of $Q_{min}$ for different CMC families. Normal bar-forming, slowly bar-forming, and bar-stable cases are presented by empty, half-filled, and filled points, respectively.}
    \label{fig:growth_rate}
\end{figure}

The bar formation timescale as a function of $Q_{min}$ and $\mathcal{C}$ can be examined in a more systematic way by fitting the growth phase of $A_{2}$ with the exponential function $a_{0}e^{\gamma t}$ where $a_{0}$ and $\gamma$ are the best-fitting variables. The latter parameter designates the exponential growth rate which is plotted in the top panel of Fig. \ref{fig:growth_rate} as a function of $Q_{min}$ for various CMC families. 
Alternatively, to verify if $\gamma$ appropriately reflects the actual timescale, the bar formation time $t_{bar}$, defined as the time at which $A_{2}$ reaches and remains above $0.2$, for all cases are plotted in the bottom panel of Fig. \ref{fig:growth_rate}.
Of interest is that although the bar-stable cases remain unbarred until $10$ Gyr, they exhibit sign of increasing $A_{2}$ that potentially attains $0.2$ beyond $10$ Gyr. To address this, the R40Q13 and R40Q15 cases are evolved further and it is found that the bars are fully formed at $10.47$ and $11.20$ Gyr, respectively, and their $\gamma$ and $t_{bar}$ are included in Fig. \ref{fig:growth_rate}. The normal bar-forming, slowly bar-forming, and bar-stable cases are differentiated by point shapes. The plot suggests that the value of $0.2\lesssim\gamma\lesssim 0.4$ can represent the slow bar formation with $8<t_{bar}<10$ Gyr, whereas those below $0.2$ yield $t_{bar}>10$ Gyr. Despite these newfound late bar formations, the classification proposed in Sec. \ref{ssec:bar_late} and \ref{ssec:kine_forming} into the normal bar-forming, the slowly bar-forming, and the bar-stable disks remain unchanged, incorporating the consideration of the proto-bar. 
It is true that the time limit of $14$ Gyr, which corresponds to the age of the Universe, was frequently employed to discriminate the bar-forming disks  \citep{fujii_et_al_2018,bland_hawthorn_et_al_2023}. We may have a different opinion that the time limit to delineate the bar-stable disks is not necessarily as long as $14$ Gyr. Because the lifetime of the proto-bar is $2$ Gyr at most according to Sec. \ref{ssec:kine_forming}, disks that have $t_{bar}$ well above $10$ Gyr are not likely to exhibit a proto-bar at the time limit of $8$ Gyr. In other words, disks with $t_{bar}>10$ Gyr can reasonably be classified as bar-stable following our scheme that is based on the dynamical state at the time limit, which results in the absence of the proto-bar. Our results so far underline the importance of the full morphological and kinematical analyses, incorporating also the cosmic time limit of $8$ Gyr, in order to classify disks into more refined categories. The morphological analysis alone is not sufficient to discriminate disks properly unless the kinematical analysis is in place.

The overall variations of $\gamma$ and $t_{bar}$ with $Q_{min}$ and $\mathcal{C}$ are in line with the earlier plots and past studies: increasing either $Q_{min}$ or $\mathcal{C}$ tends to tone down the bar growth rate \citep{mayer+wadsley_2004,berrier+sellwood_2016,kataria+das_2018,jang+kim_2023,worrakitpoonpon_2023}). The underlying mechanisms of the slowdown of the bar formation by each factor will be discussed separately. Slower bar formation for a higher $Q_{min}$ can be explained that the responsive pressure force from the velocity dispersion that counteracts the bar growth is increasing with $Q_{min}$ \citep{worrakitpoonpon_2023,cas_paper}. On the other hand, the explanation that a higher $\mathcal{C}$ can similarly slow down the bar formation can be adapted from the arguments of \citet{sellwood+evans_2001} who hypothesized that the stabilization of a disk in a concentrated halo potential against bar modes was owing to the strong shearing. Those arguments explain the presence of the long-lasting spiral traces in place of the bar in Fig. \ref{fig:snap_late} for the bar-stable case. A similar study led by \citet{bournaud_et_al_2005} conjectured the bar destruction by a high CMC via the gravity torques, also derived from the shearing. As inferred from those arguments, the formation of a rigid bar in a higher-CMC disk is subject to a stronger opposing shearing force, thus it takes a longer time for the bar modes to reach their peak. 

With a closer look, $\gamma$ decreases evidently with $Q_{min}$ in the range of $Q_{min}$ from $1.1-1.3$, whereas it apparently saturates beyond $Q_{min}=1.3$. This can be explained that disks transit from the rotation-dominated regime in which the bar modes grow rapidly and are more sensitive to subtle change of $Q_{min}$, to the pressure-dominated regime in which the velocity dispersion is sufficiently high, so the change of $Q_{min}$ does not affect much the bar formation dynamics. We note the growth rates of some Q14 disks that are higher than the Q13 counterparts, leading to lower $t_{bar}$. That tendency is absent for R75 disks as $\gamma$ monotonically decays with $Q_{min}$. This implies that the growth rate is not a monotonically decreasing function of $Q_{min}$ for some CMC values. To explain this complexity, we recall the explanations of the slowdown of the bar formation by high velocity dispersion and high shearing. It is true that increasing $Q_{min}$ results in an increase of pressure that counteracts the bar growth more efficiently but, adversely, it decreases the rotational velocity according to the Jeans equation (\ref{vthetamean}). This lessens the shearing degree which is another factor opposing the growth of the unstable bar modes.
In other words, varying $Q_{min}$ gives rise to two effects, each of which has the opposite influence to the growth rate. On the other hand, the monotonically decreasing $\gamma$ with $Q_{min}$ for the R75 family implies that the effect from increasing $Q_{min}$ to the shearing is not as important as in the other families because the concentration degree is low. That speculation is supported by the documented investigations of isolated disks without halo and bulge, which found a monotonic decrease of the growth rate with $Q_{min}$ \citep{hozumi_2022,worrakitpoonpon_2023}. Considering the effect from increasing $\mathcal{C}$, we find that $\gamma$ decreases accordingly for all $Q_{min}$. This suggests that the CMC plays a more central role in regulating the bar formation timescale than $Q_{min}$ in the explored parameter space. 

In continuity with a number of separate studies which reported the increase of the bar formation timescale by decreasing the disk fraction, equivalent to increasing the CMC \citep{fujii_et_al_2018,bland_hawthorn_et_al_2023}; increasing the CMC \citep{kataria+das_2018,jang+kim_2023}; or increasing $Q_{min}$ \citep{jang+kim_2023,cas_paper}, we demonstrate that when both parameters are varied, the growth rate is not simply a one-to-one function of $Q_{min}$ for some values of $\mathcal{C}$. In such circumstance, a warmer disk can form a bar more rapidly than a colder one. This complexity stems from the complicated roles of $Q_{min}$ and $\mathcal{C}$ in jointly overseeing the bar formation as mentioned above. 
In the observational aspect, there were reports of the anti-correlations between the bar fraction and the disk velocity dispersion \citep{sheth_et_al_2012,lee_et_al_2012,cervantes_sodi_2017} and the CMC \citep{aguerri_et_al_2009,lee+ann+park_2019}, but not many of them mentioned the possibility of the ongoing bar formation. Slow bar formation in a hot disk in the observational context was marginally mentioned by \citet{sheth_et_al_2012}. We demonstrate that delayed bar formation in a hot disk residing in a high-CMC system is possible. 

It is worth reminded that the formation timescales, especially in the isolated disk context, depend on many factors such as the galaxy physical and kinematical model, the composition, and the astrophysical processes involved. In other words, there is no absolute timescale. Our study nevertheless sheds the light on this topic that the timescales of the bar formation from a disk to another disk, if assessed by $\gamma$ or $t_{bar}$, can differ by the factor of $10$ in a carefully-controlled parameter space. 
Another remark is that we consider the bulgeless system whose applicability is limited to bulgeless galaxies. We focus on this model to avoid the complication from bulge parameters, so that the CMC is solely a function of the halo scale radius if the halo mass is fixed. With the presence of the bulge, the formation timescale could be up to $5$ times longer depending on the bulge mass and size \citep{fujii_et_al_2018,kataria+das_2018,kataria+das_2019}. Not only the bulge, the disk thickness also played a significant role such that a more vertically extended disk tended to form a bar more slowly \citep{klypin_et_al_2009,ghosh_et_al_2023,ghosh_et_al_2024}. By adding gas into a disk, it turned out that a gas-rich disk was less prone to rapid bar formation \citep{athanassoula_et_al_2013tri,seo_et_al_2019,lokas_2020gas}. Regarding the environmental effect, stochastic forces from close encounters across cosmic time were found to delay the bar emergence \citep{zana_et_al_2018}.
These factors which are not considered in our study, enhance the likeliness of slow bar formation in galaxies, which can possibly be spotted by thorough analyses on the kinematical maps proposed in Sec. \ref{ssec:kine_forming}.
We consider our work as a pilot study of implementing the time limit of $8$ Gyr to filter out the slowly bar-forming disks from the bar-stable ones, and this condition should be considered in both theoretical and observational contexts for a more appropriate classification of galaxies.

\section{Bar instability criterion revisit}
\label{sec:bar_crit}

\subsection{Is fast-rotating disk always stable?}
\label{ssec:op_eln}

\begin{figure}
    \centering
    \includegraphics[width=8.0cm]{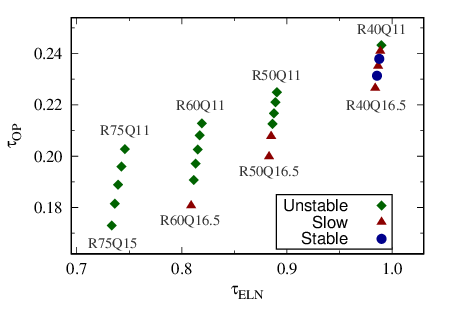}
    \caption{$\tau_{ELN}-\tau_{OP}$ diagram for all cases in Tab. \ref{tab_ini}, some of which have their names asides. Cases that are bar-stable and forming the bar before and after $8$ Gyr are differentiated by point types.}
    \label{fig:op_eln}
\end{figure}

The question of the bar instability has been a challenge for decades as a definite answer for the condition that favors the bar formation remains not fully satisfied. Notable milestones are the Ostriker-Peebles (OP) \citep{op_1973} and the Efstathiou-Lake-Negroponte (ELN) \citep*{eln_1982} criteria. The former work proposed an indicator calculated from the ratio of the rotational kinetic energy $T_{rot}$ to the magnitude of the potential energy $|U|$, namely
\begin{equation}
\tau_{OP}=\frac{T_{rot}}{|U|},
    \label{eq:op}
\end{equation}
and the disk was supposed to be stable against bar modes if $\tau_{OP}<0.14$. On the other hand, the ELN indicator is computed from the maximum tangential velocity $v_{max}$, the disk mass $M_{d}$, and the disk exponential scale radius $R_{0}$ as 
\begin{equation}
\tau_{ELN}=\frac{v_{max}}{(\frac{GM_{d}}{R_{0}})^{1/2}}.
    \label{eq:eln}
\end{equation}
The value of $\tau_{ELN}>1.1$ proved capable of stabilizing the disk according to their interpretation. The central ideas of each framework significantly differ as the OP criterion considered the partition of the kinetic energies in the global level, whereas the ELN counterpart focused on fine details of the rotational curve, namely the local consideration. To inspect if these criteria conform with our result, $\tau_{ELN}$ and $\tau_{OP}$ for all cases in Tab. \ref{tab_ini} are plotted in Fig. \ref{fig:op_eln}. We consider the slowly bar-forming disks as unbarred like the bar-stable cases due to limited evolution time, bearing in mind that they are dynamically distinct, and they morphologically differ from the normal bar-forming ones which become barred within the time limit.

The OP criterion provides acceptable description of the overall disk stability as the normal bar-forming disks are placed above $0.14$. However, we have two concerns on the OP scope. Firstly, the rotationally-dominated disks, designated by high $\tau_{OP}$, are not necessarily bar-unstable as proved by the R40 family which is situated well above $0.14$ but most of the members are either slowly bar-forming or bar-stable. The original calculation of the OP index was on the basis that a lower fraction gave rise to higher counter-acting pressure to suppress the bar growth, as suggested by numerous perturbative analyses \citep{christodoulou_et_al_1995,khoperskov_et_al_2003,jalali+hunter_2005,polyachenko+just_2015}. This contradicts our results at the high-$\tau_{OP}$ end where the fast-rotating R40 family is located. As discussed above, the stabilization is attributed to the strong shearing by high $\mathcal{C}$. These suggest that the applicability of the OP criterion is limited to cases without strong shearing. The exception is when $Q_{min}$ is low so that the shearing is unable to neutralize the fast-growing bar modes, which is the situation for R40Q11.
Secondly, the critical $\tau_{OP}$ is possibly $\mathcal{C}$-dependent. For instance, the critical $\tau_{OP}=0.14$ is applicable to the R75 family only, whereas it has to be $0.18$, $0.21$ and $0.24$ to delineate normal bar-forming disks above for R60, R50, and R40 families, respectively.

Considering the ELN criterion, it turns out that in order to cope with all bar-stable cases and a good fraction of slowly bar-forming cases, the critical $\tau_{ELN}$ has to be slightly shifted down to $\sim 1$. The discrepancy between the original ELN threshold and our result can be explained that the ELN indicator depends on the fine details of the rotation curve, which is sensitive to the choices of the configurational, compositional and kinematical models of the initial disk. Since our model does not consist of a bulge, the disk center rotates more slowly for a given $Q_{min}$, leading to a lower critical $\tau_{ELN}$. Generally speaking, the critical $\tau_{ELN}$ is model-dependent.
If we neglect the slight mismatch between the original and our critical $\tau_{ELN}$, the ELN framework provides a better description of the disk stability than the OP counterpart as disks with $\tau_{ELN}$ exceeding the critical value are the fast-rotating ones, yielding a higher shearing degree which is the major factor in stabilizing the disks against bar modes according to our finding. That the CMC plays a more significant role than $Q_{min}$ can also be seen in the diagram as varying $Q_{min}$ in each CMC family alters $\tau_{ELN}$ slightly compared to varying $\mathcal{C}$. However, the case with highest $\tau_{ELN}$, namely R40Q11, is still prone to the bar instability. This underlines the necessity of high $Q_{min}$ along with high $\mathcal{C}$ for the disk stabilization. That the supercritical-$\tau_{ELN}$ disk is not necessarily bar-stable is in accordance with observational works that examined the bar instability in disk galaxies of various properties \citep{athanassoula_2008let,ghosh_et_al_2023,romeo_et_al_2023}. It was found that fast-rotating disks turned to be barred and unbarred in comparable fractions.

The disk stabilization by a high CMC was also concluded by \citet{saha+elmegreen_2018} who explored the bar instability in the systems of various bulge concentrations. Such model yielded the prominent peak of $\Omega-\kappa/2$ in the disk interior if a compact bulge was present. As a consequence, particles close to the disk center could be trapped by the bi-symmetric potential owing to the inner Lindblad resonance which prevented bar growth. Compared to our study, the $\Omega-\kappa/2$ curves do not exhibit prominent peaks (see Fig. \ref{fig:fou_spec}), but a high CMC value can similarly suppress the bar formation. We incline towards the strong shearing as the major stabilization factor, and the inner Lindblad resonance plays a less important role as it does not occur in our cases but the disks can nevertheless be stable. The persistent spiral trace is the proof of the strong shearing.

\subsection{Stability analysis on the $Q_{min}$-CMC diagram}
\label{ssec:q_cmc_rev}

\begin{figure}
    \centering
    \includegraphics[width=8.0cm]{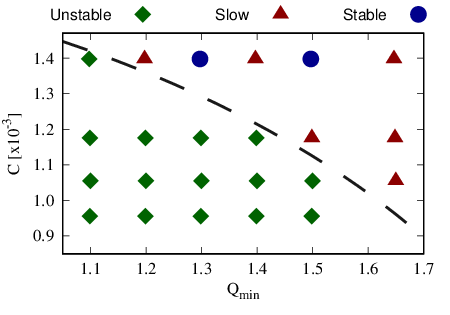}
    \caption{$Q_{min}$-CMC diagram for cases listed in Tab. \ref{tab_ini}. The bar-stable cases and the cases than form a bar before and after $8$ Gyr are distinguished by point shapes. The dashed line represents the relation $\frac{Q_{min}^{2}}{4}+\frac{\mathcal{C}^{2}}{2.89}=1$ which partitions the normal bar-forming case from the two other groups.}
    \label{fig:qmin_cmc}
\end{figure}

We revisit the stability criterion involving $Q_{min}$ and CMC, both of which have been conjectured to be the underlying factors for the growth rate and the stabilization. We follow the $Q_{min}$-CMC diagram proposed by \citet{jang+kim_2023} who were able to identify the stable and unstable regions in such diagram, and our corresponding $Q_{min}$-CMC diagram is shown in Fig. \ref{fig:qmin_cmc}. We distribute cases more evenly in the $Q_{min}$-CMC diagram so that the partition between regimes can be clearly identified. The cases that are bar-stable and the cases that form a bar before and after $8$ Gyr are differentiated by point shapes. Considering the normal bar-forming cases, they can be enclosed by the elliptical equation 
\begin{equation}
\frac{Q_{min}^{2}}{4}+\frac{\mathcal{C}^{2}}{2.89}=1
\label{eq:ellip}
\end{equation}
in accordance with the finding of \citet{jang+kim_2023}, although we exploit a different disk model. However, cases above the delineating line deserve attention. First of all, the bar-stable and the slowly bar-forming cases cannot be separated by a simple one-to-one relation as that enclosing the normal bar-unstable cases beneath. This is attributed to the complicated interplay between $Q_{min}$ and $\mathcal{C}$, as mentioned in Sec. \ref{ssec:growth_rate}, that makes the growth rate not monotonically decrease with $Q_{min}$ in the high-$\mathcal{C}$ regime. 
Another remark from the $Q_{min}$-CMC diagram is that a high $Q_{min}$, even close to the limit for the physical solution of the Jeans equation to exist, cannot suppress the bar formation unless $\mathcal{C}$ reaches the level for the halo of $r_{h}=40$ kpc. For the CMC level below, the diagram suggests the prolongation of the bar formation process to the point that it can be misidentified as bar-stable if it is not evolved long enough. 
Possible explanation is that the growth rate of the global unstable bar modes depends on the perturbation pattern speed in an increasing way, as inferred from some theoretical and numerical studies \citep{jalali_2007,polyachenko+just_2015}. Thus, the definite suppression of the bar growth without a high $\mathcal{C}$ is possible only when the disk is non-rotating, which is not the realistic case.

That a high $\mathcal{C}$ is required asides a high $Q_{min}$ to effectively suppress the bar instability contradicts past stability analyses on the isolated disk, proving that the global unstable two-armed modes could be subdued by a certain degree of the velocity dispersion \citep{vandervoort_1982,lemos_et_al_1991,christodoulou_et_al_1995,khoperskov_et_al_2003,jalali+hunter_2005}. In other words, a slowly rotating disk is more likely to be bar-stable. The discrepancy is attributed to the presence of the bulge or the halo which causes the disk to systematically rotate faster than being in isolation. This has a destabilizing effect on the disk so that it is no longer possible to subdue the bar growth only by high velocity dispersion. 
Likewise, a high CMC alone cannot stabilize the disk as seen in the R40Q11 case which possesses highest $\tau_{ELN}$ but it manifests normal bar formation. 
However, some R40 cases with $Q_{min}>1.2$ are effectively stabilized. To conclude this part, in order to stabilize the disk efficiently, both high $Q_{min}$ and CMC are necessary as a sufficiently high $Q_{min}$ is needed to attenuate the growth of the global unstable bar modes so that they can be neutralized by the shearing. 
Let us discuss the involvement of the swing amplification. Our finding suggests that such mechanism plays different roles in each regime. It was documented that the swing amplification promoted bar formation by means of introducing the non-axisymmetric modes in addition to the unstable two-armed modes. This led to the systematic elevation of the critical $Q$ when transiting from the rigidly rotating to the differentially rotating disks \citep{worrakitpoonpon_2023}. The arising of the swing amplification might be the explanation of the unavoidable instability even though $Q_{min}$ is close to the limit in the low-CMC regime. On the contrary, the role of the swing amplification differs when $\mathcal{C}$ is sufficiently high. In conjunction with a high $Q_{min}$, the swing amplification rather gives rise to the long-lasting spiral structure, owing to the weakened bar modes, to the point that the disk can be classified as bar-stable due to the absence of the proto-bar. However, a low $Q_{min}$ can render normal bar formation, albeit high $\mathcal{C}$, because the shearing is not able to overtake the fast-growing global bar modes and the bar modes overwhelm the spiral modes in short time.

The presence of slowly bar-forming disks that are currently unbarred, if identifiable, might significantly alter the galaxy census from the surveys as our results suggested that disk galaxies possessing either high $Q_{min}$ or high $\mathcal{C}$ are not necessarily bar-stable.  
Current galaxy classification is based on the morphological, physical, or photometric parameters only such as the bar strength, the concentration, the asymmetry, or the color index. Therefore, the slowly bar-forming and the stable disks are inevitably categorized as unbarred, as seen by subtle difference between the bar amplitudes and the configurations in the middle of evolution (see Fig. \ref{fig:a2} and \ref{fig:snap_late}). We demonstrate that with some complementary kinematical maps, the two sub-groups in the unbarred category can be appropriately distinguished.  
It is worth noting that our $Q_{min}$-CMC diagram is specific to the bulgeless model in which $\mathcal{C}$ is simply a function of the halo scale radius if the halo mass is fixed. Thus, the two-parameter diagram is sufficient to represent the stability condition. A disk model including a bulge gives rise to a more complicated diagram that involves more than $2$ parameters. This, however, does not hinder the concepts of the slow bar formation and the proto-bar as the presence of the bulge was found to prolong the bar formation process compared to the bulgeless system. This renders greater possibility of the existence of the slowly bar-forming galaxies among the observed ones.

\section{Conclusion}
\label{sec:concls}

We used $N$-body simulations to investigate the timescale of the process of the bar formation and the disk stability across the range of the Toomre's parameter $Q_{min}$ and the central mass concentration (CMC) parameter, namely $\mathcal{C}$. Apart from the elementary finding that the timescales could range from $1-10$ Gyr depending on those parameters, we proposed the limit of the time in which the bar was allowed to develop secularly in real galaxies, prescribed by the Universe evolution timeline. More specifically, according to recent concordance cosmological simulations, a secular bar growth in a typical disk galaxy could not be initiated earlier than $8$ Gyr of the lookback time, on average, at which the disk became kinematically and compositionally steady to prompt the bar instability. In our consideration, disks that became fully barred after this time limit were classified as the slowly bar-forming disks. 
In the $Q_{min}$-CMC diagram, such slowly bar-forming disks manifested in the range of intermediate to high $Q_{min}$ and $\mathcal{C}$ well above those that formed a bar normally, i.e., before $8$ Gyr, and just below the bar-stable region, populated by disks that remained unbarred until at least $10$ Gyr.
The latter group occupied the highest end of that diagram. The same diagram suggested that the truly bar-stable disks could be obtained by both high $Q_{min}$ and high $\mathcal{C}$. Either elevated $Q_{min}$ or $\mathcal{C}$ did not suffice as it only prolonged the bar formation process.

The other concern was that the slowly bar-forming disks were nearly indistinguishable from the stable disks prior to $8$ Gyr, if evaluated by the bar amplitudes and the configurations. 
To address this, we proposed the complementary methods to differentiate between the two groups. We found that by inspecting the Fourier spectrogram, the fine-scale surface density map, and the angular frequency field, we could effectively filter out these slowly bar-forming disks from the bar-stable ones.
The former group had the proto-bar embedding in the disk center, which exhibited features specific to the accumulation of particles forming a rigid bar as seen in the three plots. The proto-bar emerged well before $8$ Gyr when the bar strength was as low as $0.1$ and it could be spotted at as early as $2$ Gyr before the fully developed bar. 
On the other hand, the proto-bar was absent in the bar-stable disk until $8$ Gyr. Although some extended runs eventually became barred beyond $10$ Gyr, the fact that the lifetime of the proto-bar was $2$ Gyr at most suggested that disks that formed a bar after $10$ Gyr could still reasonably be classified as bar-stable by the absence of the proto-bar at $8$ Gyr. In summary, our full morphological, kinematical, and temporal analyses allowed us to better classify disks into more refined categories.

The concept of the proto-bar can be applied to the observational context as it helps to identify the bar-forming galaxies which do not yet possess a visible bar at present, if the proto-bar stage coincides with the time of the observation and the telescope resolution permits to do so. 
The analysis on the kinematical maps in search of the proto-bar is complementary to the conventional morphological analysis on the surface brightness, i.e., the photometric analysis, as the kinematical maps are based on the spectroscopic data.
Their existences lead to the reconsideration of the galaxy census as the slowly bar-forming and the stable disks are dynamically distinct. This implies that the bar-stable disks constitute a lower fraction than previously reported. 

In addition, we found that the bar growth rate did not decrease monotonically with $Q_{min}$ for some values of the CMC. More specifically, some cases with $Q_{min}=1.4$ developed a bar more rapidly than the $Q_{min}=1.3$ counterparts. It was because increasing $Q_{min}$ led to two opposite effects on the bar growth. On the one hand, increasing $Q_{min}$ raised the pressure force that counteracted the bar growth more efficiently. On the other hand, doing that reduced the shearing degree which had the neutralizing effect to the bar modes. This implied that the interplays between $Q_{min}$ and the CMC in regulating the bar instability and the bar growth were more complicated than previously understood.

\begin{acknowledgements}

  This work is supported by (i) Suranaree University of Technology (SUT), (ii) Thailand Science Research and Innovation (TSRI), and (iii) National Science, Research and Innovation Fund (NSRF), project no. 195242, and partly by the Program Management Unit for Human Resources \& Institutional Development, Research and Innovation (PMUB), grant number B16F640076. Numerical simulations are facilitated by HPC resources of Chalawan cluster of the National Astronomical Research Institute of Thailand.

\end{acknowledgements}


\begin{thebibliography}{}
\expandafter\ifx\csname natexlab\endcsname\relax\def\natexlab#1{#1}\fi
\providecommand{\url}[1]{\href{#1}{#1}}
\providecommand{\dodoi}[1]{doi:~\href{http://doi.org/#1}{\nolinkurl{#1}}}
\providecommand{\doeprint}[1]{\href{http://ascl.net/#1}{\nolinkurl{http://ascl.net/#1}}}
\providecommand{\doarXiv}[1]{\href{https://arxiv.org/abs/#1}{\nolinkurl{https://arxiv.org/abs/#1}}}

\bibitem[{{Aguerri} {et~al.}(2009){Aguerri}, {M{\'e}ndez-Abreu}, \&
  {Corsini}}]{aguerri_et_al_2009}
{Aguerri}, J.~A.~L., {M{\'e}ndez-Abreu}, J., \& {Corsini}, E.~M. 2009, \aap,
  495, 491, \dodoi{10.1051/0004-6361:200810931}

\bibitem[{{Aguerri} {et~al.}(2015){Aguerri}, {M{\'e}ndez-Abreu},
  {Falc{\'o}n-Barroso}, {Amorin}, {Barrera-Ballesteros}, {Cid Fernandes},
  {Garc{\'\i}a-Benito}, {Garc{\'\i}a-Lorenzo}, {Gonz{\'a}lez Delgado},
  {Husemann}, {Kalinova}, {Lyubenova}, {Marino}, {M{\'a}rquez}, {Mast},
  {P{\'e}rez}, {S{\'a}nchez}, {van de Ven}, {Walcher}, {Backsmann},
  {Cortijo-Ferrero}, {Bland-Hawthorn}, {del Olmo}, {Iglesias-P{\'a}ramo},
  {P{\'e}rez}, {S{\'a}nchez-Bl{\'a}zquez}, {Wisotzki}, \&
  {Ziegler}}]{aguerri_et_al_2015}
{Aguerri}, J.~A.~L., {M{\'e}ndez-Abreu}, J., {Falc{\'o}n-Barroso}, J., {et~al.}
  2015, \aap, 576, A102, \dodoi{10.1051/0004-6361/201423383}

\bibitem[{{Algorry} {et~al.}(2017){Algorry}, {Navarro}, {Abadi}, {Sales},
  {Bower}, {Crain}, {Dalla Vecchia}, {Frenk}, {Schaller}, {Schaye}, \&
  {Theuns}}]{algorry_et_al_2017}
{Algorry}, D.~G., {Navarro}, J.~F., {Abadi}, M.~G., {et~al.} 2017, \mnras, 469,
  1054, \dodoi{10.1093/mnras/stx1008}

\bibitem[{{Ardi} {et~al.}(2003){Ardi}, {Tsuchiya}, \&
  {Burkert}}]{ardi_et_al_2003}
{Ardi}, E., {Tsuchiya}, T., \& {Burkert}, A. 2003, \apj, 596, 204,
  \dodoi{10.1086/377684}

\bibitem[{{Athanassoula}(2003)}]{athanassoula_2003}
{Athanassoula}, E. 2003, \mnras, 341, 1179,
  \dodoi{10.1046/j.1365-8711.2003.06473.x}

\bibitem[{{Athanassoula}(2008)}]{athanassoula_2008let}
---. 2008, \mnras, 390, L69, \dodoi{10.1111/j.1745-3933.2008.00541.x}

\bibitem[{{Athanassoula} {et~al.}(2005){Athanassoula}, {Lambert}, \&
  {Dehnen}}]{athanassoula_et_al_2005}
{Athanassoula}, E., {Lambert}, J.~C., \& {Dehnen}, W. 2005, \mnras, 363, 496,
  \dodoi{10.1111/j.1365-2966.2005.09445.x}

\bibitem[{{Athanassoula} {et~al.}(2013){Athanassoula}, {Machado}, \&
  {Rodionov}}]{athanassoula_et_al_2013tri}
{Athanassoula}, E., {Machado}, R. E.~G., \& {Rodionov}, S.~A. 2013, \mnras,
  429, 1949, \dodoi{10.1093/mnras/sts452}

\bibitem[{{Baba} \& {Kawata}(2020)}]{baba+kawata_2020}
{Baba}, J., \& {Kawata}, D. 2020, \mnras, 492, 4500,
  \dodoi{10.1093/mnras/staa140}

\bibitem[{{Barazza} {et~al.}(2009){Barazza}, {Jablonka}, {Desai}, {Jogee},
  {Arag{\'o}n-Salamanca}, {De Lucia}, {Saglia}, {Halliday}, {Poggianti},
  {Dalcanton}, {Rudnick}, {Milvang-Jensen}, {Noll}, {Simard}, {Clowe},
  {Pell{\'o}}, {White}, \& {Zaritsky}}]{barazza_et_al_2009}
{Barazza}, F.~D., {Jablonka}, P., {Desai}, V., {et~al.} 2009, \aap, 497, 713,
  \dodoi{10.1051/0004-6361/200810352}

\bibitem[{{Barway} {et~al.}(2016){Barway}, {Saha}, {Vaghmare}, \&
  {Kembhavi}}]{barway_et_al_2016}
{Barway}, S., {Saha}, K., {Vaghmare}, K., \& {Kembhavi}, A.~K. 2016, \mnras,
  463, L41, \dodoi{10.1093/mnrasl/slw153}

\bibitem[{{Bauer} \& {Widrow}(2019)}]{bauer+widrow_2019}
{Bauer}, J.~S., \& {Widrow}, L.~M. 2019, \mnras, 486, 523,
  \dodoi{10.1093/mnras/stz478}

\bibitem[{{Berentzen} {et~al.}(2006){Berentzen}, {Shlosman}, \&
  {Jogee}}]{berentzen_et_al_2006}
{Berentzen}, I., {Shlosman}, I., \& {Jogee}, S. 2006, \apj, 637, 582,
  \dodoi{10.1086/498493}

\bibitem[{{Berrier} \& {Sellwood}(2016)}]{berrier+sellwood_2016}
{Berrier}, J.~C., \& {Sellwood}, J.~A. 2016, \apj, 831, 65,
  \dodoi{10.3847/0004-637X/831/1/65}

\bibitem[{{Bland-Hawthorn} {et~al.}(2023){Bland-Hawthorn}, {Tepper-Garcia},
  {Agertz}, \& {Freeman}}]{bland_hawthorn_et_al_2023}
{Bland-Hawthorn}, J., {Tepper-Garcia}, T., {Agertz}, O., \& {Freeman}, K. 2023,
  \apj, 947, 80, \dodoi{10.3847/1538-4357/acc469}

\bibitem[{{Bournaud} {et~al.}(2005){Bournaud}, {Combes}, {Jog}, \&
  {Puerari}}]{bournaud_et_al_2005}
{Bournaud}, F., {Combes}, F., {Jog}, C.~J., \& {Puerari}, I. 2005, \aap, 438,
  507, \dodoi{10.1051/0004-6361:20052631}

\bibitem[{{Buta} {et~al.}(2019){Buta}, {Verdes-Montenegro}, {Damas-Segovia},
  {Jones}, {Blasco}, {Fern{\'a}ndez-Lorenzo}, {Sanchez}, {Garrido},
  {Ramirez-Moreta}, \& {Sulentic}}]{buta_et_al_2019}
{Buta}, R.~J., {Verdes-Montenegro}, L., {Damas-Segovia}, A., {et~al.} 2019,
  \mnras, 488, 2175, \dodoi{10.1093/mnras/stz1780}

\bibitem[{{Cervantes Sodi}(2017)}]{cervantes_sodi_2017}
{Cervantes Sodi}, B. 2017, \apj, 835, 80, \dodoi{10.3847/1538-4357/835/1/80}

\bibitem[{{Chantavat} {et~al.}(2024){Chantavat}, {Yuma}, {Malelohit}, \&
  {Worrakitpoonpon}}]{cas_paper}
{Chantavat}, T., {Yuma}, S., {Malelohit}, P., \& {Worrakitpoonpon}, T. 2024,
  \apj, 965, 77, \dodoi{10.3847/1538-4357/ad3218}

\bibitem[{{Chiba} \& {Sch{\"o}nrich}(2022)}]{chiba+schonrich_2022}
{Chiba}, R., \& {Sch{\"o}nrich}, R. 2022, \mnras, 513, 768,
  \dodoi{10.1093/mnras/stac697}

\bibitem[{{Christodoulou} {et~al.}(1995){Christodoulou}, {Shlosman}, \&
  {Tohline}}]{christodoulou_et_al_1995}
{Christodoulou}, D.~M., {Shlosman}, I., \& {Tohline}, J.~E. 1995, \apj, 443,
  551, \dodoi{10.1086/175547}

\bibitem[{{Cole} \& {Weinberg}(2002)}]{cole+weinberg_2002}
{Cole}, A.~A., \& {Weinberg}, M.~D. 2002, \apjl, 574, L43,
  \dodoi{10.1086/342278}

\bibitem[{{Collier} {et~al.}(2019){Collier}, {Shlosman}, \&
  {Heller}}]{collier_et_al_2019b}
{Collier}, A., {Shlosman}, I., \& {Heller}, C. 2019, \mnras, 489, 3102,
  \dodoi{10.1093/mnras/stz2327}

\bibitem[{{Combes} \& {Sanders}(1981)}]{combes+sanders_1981}
{Combes}, F., \& {Sanders}, R.~H. 1981, \aap, 96, 164

\bibitem[{{Costantin} {et~al.}(2023){Costantin}, {P{\'e}rez-Gonz{\'a}lez},
  {Guo}, {Buttitta}, {Jogee}, {Bagley}, {Barro}, {Kartaltepe}, {Koekemoer},
  {Cabello}, {Corsini}, {M{\'e}ndez-Abreu}, {de la Vega}, {Iyer}, {Bisigello},
  {Cheng}, {Morelli}, {Arrabal Haro}, {Buitrago}, {Cooper}, {Dekel},
  {Dickinson}, {Finkelstein}, {Giavalisco}, {Holwerda}, {Huertas-Company},
  {Lucas}, {Papovich}, {Pirzkal}, {Seill{\'e}}, {Vega-Ferrero}, {Wuyts}, \&
  {Yung}}]{constantin_et_al_2023}
{Costantin}, L., {P{\'e}rez-Gonz{\'a}lez}, P.~G., {Guo}, Y., {et~al.} 2023,
  \nat, 623, 499, \dodoi{10.1038/s41586-023-06636-x}

\bibitem[{{Cuomo} {et~al.}(2024){Cuomo}, {Morelli}, {Aguerri}, {Corsini},
  {Debattista}, {Coccato}, {Pizzella}, {Boselli}, {Buttitta}, {de
  Lorenzo-C{\'a}ceres}, {Ferrarese}, {Gasparri}, {Lee}, {Mendez-Abreu},
  {Roediger}, \& {Zarattini}}]{cuomo_et_al_2024}
{Cuomo}, V., {Morelli}, L., {Aguerri}, J. A.~L., {et~al.} 2024, \mnras, 527,
  11218, \dodoi{10.1093/mnras/stad3945}

\bibitem[{{De Rijcke} {et~al.}(2019){De Rijcke}, {Fouvry}, \&
  {Pichon}}]{de_rijcke_et_al_2019}
{De Rijcke}, S., {Fouvry}, J.-B., \& {Pichon}, C. 2019, \mnras, 484, 3198,
  \dodoi{10.1093/mnras/stz166}

\bibitem[{{de S{\'a}-Freitas} {et~al.}(2023){de S{\'a}-Freitas}, {Fragkoudi},
  {Gadotti}, {Falc{\'o}n-Barroso}, {Bittner}, {S{\'a}nchez-Bl{\'a}zquez}, {van
  de Ven}, {Bieri}, {Coccato}, {Coelho}, {Fahrion}, {Gon{\c{c}}alves}, {Kim},
  {de Lorenzo-C{\'a}ceres}, {Martig}, {Mart{\'\i}n-Navarro}, {Mendez-Abreu},
  {Neumann}, \& {Querejeta}}]{de_sa_freitas_et_al_2023}
{de S{\'a}-Freitas}, C., {Fragkoudi}, F., {Gadotti}, D.~A., {et~al.} 2023,
  \aap, 671, A8, \dodoi{10.1051/0004-6361/202244667}

\bibitem[{{Dubinski} {et~al.}(2009){Dubinski}, {Berentzen}, \&
  {Shlosman}}]{dubinski_et_al_2009}
{Dubinski}, J., {Berentzen}, I., \& {Shlosman}, I. 2009, \apj, 697, 293,
  \dodoi{10.1088/0004-637X/697/1/293}

\bibitem[{{Efstathiou} {et~al.}(1982){Efstathiou}, {Lake}, \&
  {Negroponte}}]{eln_1982}
{Efstathiou}, G., {Lake}, G., \& {Negroponte}, J. 1982, \mnras, 199, 1069,
  \dodoi{10.1093/mnras/199.4.1069}

\bibitem[{{Erwin}(2018)}]{erwin_2018}
{Erwin}, P. 2018, \mnras, 474, 5372, \dodoi{10.1093/mnras/stx3117}

\bibitem[{{Eskridge} {et~al.}(2000){Eskridge}, {Frogel}, {Pogge}, {Quillen},
  {Davies}, {DePoy}, {Houdashelt}, {Kuchinski}, {Ram{\'\i}rez}, {Sellgren},
  {Terndrup}, \& {Tiede}}]{eskridge_et_al_2000}
{Eskridge}, P.~B., {Frogel}, J.~A., {Pogge}, R.~W., {et~al.} 2000, \aj, 119,
  536, \dodoi{10.1086/301203}

\bibitem[{{Frosst} {et~al.}(2024){Frosst}, {Obreschkow}, \&
  {Ludlow}}]{frosst_et_al_2024}
{Frosst}, M., {Obreschkow}, D., \& {Ludlow}, A. 2024, \mnras, 534, 313,
  \dodoi{10.1093/mnras/stae2086}

\bibitem[{{Fujii} {et~al.}(2018){Fujii}, {B{\'e}dorf}, {Baba}, \& {Portegies
  Zwart}}]{fujii_et_al_2018}
{Fujii}, M.~S., {B{\'e}dorf}, J., {Baba}, J., \& {Portegies Zwart}, S. 2018,
  \mnras, 477, 1451, \dodoi{10.1093/mnras/sty711}

\bibitem[{{Fujimoto} {et~al.}(2023){Fujimoto}, {Inutsuka}, \&
  {Baba}}]{fujimoto_et_al_2023}
{Fujimoto}, Y., {Inutsuka}, S.-i., \& {Baba}, J. 2023, \mnras, 523, 3049,
  \dodoi{10.1093/mnras/stad1612}

\bibitem[{{G{\'e}ron} {et~al.}(2023){G{\'e}ron}, {Smethurst}, {Lintott},
  {Kruk}, {Masters}, {Simmons}, {Mantha}, {Walmsley}, {Garma-Oehmichen},
  {Drory}, \& {Lane}}]{geron_et_al_2023}
{G{\'e}ron}, T., {Smethurst}, R.~J., {Lintott}, C., {et~al.} 2023, \mnras, 521,
  1775, \dodoi{10.1093/mnras/stad501}

\bibitem[{{Ghosh} {et~al.}(2023){Ghosh}, {Fragkoudi}, {Di Matteo}, \&
  {Saha}}]{ghosh_et_al_2023}
{Ghosh}, S., {Fragkoudi}, F., {Di Matteo}, P., \& {Saha}, K. 2023, \aap, 674,
  A128, \dodoi{10.1051/0004-6361/202245275}

\bibitem[{{Ghosh} {et~al.}(2024){Ghosh}, {Gadotti}, {Fragkoudi}, {Nagpal}, {Di
  Matteo}, \& {Cuomo}}]{ghosh_et_al_2024}
{Ghosh}, S., {Gadotti}, D.~A., {Fragkoudi}, F., {et~al.} 2024, \mnras, 532,
  4570, \dodoi{10.1093/mnras/stae1797}

\bibitem[{{Grand} {et~al.}(2017){Grand}, {G{\'o}mez}, {Marinacci}, {Pakmor},
  {Springel}, {Campbell}, {Frenk}, {Jenkins}, \& {White}}]{grand_et_al_2017}
{Grand}, R. J.~J., {G{\'o}mez}, F.~A., {Marinacci}, F., {et~al.} 2017, \mnras,
  467, 179, \dodoi{10.1093/mnras/stx071}

\bibitem[{{Guo} {et~al.}(2023){Guo}, {Jogee}, {Finkelstein}, {Chen}, {Wise},
  {Bagley}, {Barro}, {Wuyts}, {Kocevski}, {Kartaltepe}, {McGrath}, {Ferguson},
  {Mobasher}, {Giavalisco}, {Lucas}, {Zavala}, {Lotz}, {Grogin},
  {Huertas-Company}, {Vega-Ferrero}, {Hathi}, {Arrabal Haro}, {Dickinson},
  {Koekemoer}, {Papovich}, {Pirzkal}, {Yung}, {Backhaus}, {Bell},
  {Calabr{\`o}}, {Cleri}, {Coogan}, {Cooper}, {Costantin}, {Croton}, {Davis},
  {Dekel}, {Franco}, {Gardner}, {Holwerda}, {Hutchison}, {Pandya},
  {P{\'e}rez-Gonz{\'a}lez}, {Ravindranath}, {Rose}, {Trump}, {de la Vega}, \&
  {Wang}}]{guo_et_al_2023barjwst}
{Guo}, Y., {Jogee}, S., {Finkelstein}, S.~L., {et~al.} 2023, \apjl, 945, L10,
  \dodoi{10.3847/2041-8213/acacfb}

\bibitem[{{Gustafsson} {et~al.}(2016){Gustafsson}, {Church}, {Davies}, \&
  {Rickman}}]{gustafsson_et_al_2016}
{Gustafsson}, B., {Church}, R.~P., {Davies}, M.~B., \& {Rickman}, H. 2016,
  \aap, 593, A85, \dodoi{10.1051/0004-6361/201423916}

\bibitem[{{H{\"a}nninen} \& {Flynn}(2002)}]{hanninen+flynn_2002}
{H{\"a}nninen}, J., \& {Flynn}, C. 2002, \mnras, 337, 731,
  \dodoi{10.1046/j.1365-8711.2002.05956.x}

\bibitem[{{Hernquist}(1990)}]{hernquist_1990}
{Hernquist}, L. 1990, \apj, 356, 359, \dodoi{10.1086/168845}

\bibitem[{{Hernquist}(1993)}]{hernquist_1993}
---. 1993, \apjs, 86, 389, \dodoi{10.1086/191784}

\bibitem[{{Hohl}(1971)}]{hohl_1971}
{Hohl}, F. 1971, \apj, 168, 343, \dodoi{10.1086/151091}

\bibitem[{{Hohl}(1976)}]{hohl_1976}
---. 1976, \aj, 81, 30, \dodoi{10.1086/111849}

\bibitem[{{Holley-Bockelmann} {et~al.}(2005){Holley-Bockelmann}, {Weinberg}, \&
  {Katz}}]{holley-bockelmann_et_al_2005}
{Holley-Bockelmann}, K., {Weinberg}, M., \& {Katz}, N. 2005, \mnras, 363, 991,
  \dodoi{10.1111/j.1365-2966.2005.09501.x}

\bibitem[{{Hozumi}(2022)}]{hozumi_2022}
{Hozumi}, S. 2022, \mnras, 510, 4394, \dodoi{10.1093/mnras/stab3704}

\bibitem[{{Izquierdo-Villalba} {et~al.}(2022){Izquierdo-Villalba}, {Bonoli},
  {Rosas-Guevara}, {Springel}, {White}, {Zana}, {Dotti}, {Spinoso}, {Bonetti},
  \& {Lupi}}]{izquierdo_villalba_et_al_2022}
{Izquierdo-Villalba}, D., {Bonoli}, S., {Rosas-Guevara}, Y., {et~al.} 2022,
  \mnras, 514, 1006, \dodoi{10.1093/mnras/stac1413}

\bibitem[{{Jackson} {et~al.}(2020){Jackson}, {Martin}, {Kaviraj}, {Laigle},
  {Devriendt}, {Dubois}, \& {Pichon}}]{jackson_et_al_2020a}
{Jackson}, R.~A., {Martin}, G., {Kaviraj}, S., {et~al.} 2020, \mnras, 494,
  5568, \dodoi{10.1093/mnras/staa970}

\bibitem[{{Jalali}(2007)}]{jalali_2007}
{Jalali}, M.~A. 2007, \apj, 669, 218, \dodoi{10.1086/521523}

\bibitem[{{Jalali} \& {Hunter}(2005)}]{jalali+hunter_2005}
{Jalali}, M.~A., \& {Hunter}, C. 2005, \apj, 630, 804, \dodoi{10.1086/432370}

\bibitem[{{James} \& {Percival}(2016)}]{james+percival_2016}
{James}, P.~A., \& {Percival}, S.~M. 2016, \mnras, 457, 917,
  \dodoi{10.1093/mnras/stv2978}

\bibitem[{{Jang} \& {Kim}(2023)}]{jang+kim_2023}
{Jang}, D., \& {Kim}, W.-T. 2023, \apj, 942, 106,
  \dodoi{10.3847/1538-4357/aca7bc}

\bibitem[{{Jenkins} \& {Binney}(1990)}]{jenkins+binney_1990}
{Jenkins}, A., \& {Binney}, J. 1990, \mnras, 245, 305

\bibitem[{{Jogee} {et~al.}(2004){Jogee}, {Barazza}, {Rix}, {Shlosman},
  {Barden}, {Wolf}, {Davies}, {Heyer}, {Beckwith}, {Bell}, {Borch}, {Caldwell},
  {Conselice}, {Dahlen}, {H{\"a}ussler}, {Heymans}, {Jahnke}, {Knapen},
  {Laine}, {Lubell}, {Mobasher}, {McIntosh}, {Meisenheimer}, {Peng},
  {Ravindranath}, {Sanchez}, {Somerville}, \& {Wisotzki}}]{jogee_et_al_2004}
{Jogee}, S., {Barazza}, F.~D., {Rix}, H.-W., {et~al.} 2004, \apjl, 615, L105,
  \dodoi{10.1086/426138}

\bibitem[{{Joshi} \& {Widrow}(2024)}]{joshi+widrow_2024}
{Joshi}, R., \& {Widrow}, L.~M. 2024, \mnras, 527, 7781,
  \dodoi{10.1093/mnras/stad3666}

\bibitem[{{Julian} \& {Toomre}(1966)}]{julian+toomre_1966}
{Julian}, W.~H., \& {Toomre}, A. 1966, \apj, 146, 810, \dodoi{10.1086/148957}

\bibitem[{{Kataria} \& {Das}(2018)}]{kataria+das_2018}
{Kataria}, S.~K., \& {Das}, M. 2018, \mnras, 475, 1653,
  \dodoi{10.1093/mnras/stx3279}

\bibitem[{{Kataria} \& {Das}(2019)}]{kataria+das_2019}
---. 2019, \apj, 886, 43, \dodoi{10.3847/1538-4357/ab48f7}

\bibitem[{{Khoperskov} {et~al.}(2003){Khoperskov}, {Zasov}, \&
  {Tyurina}}]{khoperskov_et_al_2003}
{Khoperskov}, A.~V., {Zasov}, A.~V., \& {Tyurina}, N.~V. 2003, Astronomy
  Reports, 47, 357, \dodoi{10.1134/1.1575851}

\bibitem[{{Klypin} {et~al.}(2009){Klypin}, {Valenzuela}, {Col{\'\i}n}, \&
  {Quinn}}]{klypin_et_al_2009}
{Klypin}, A., {Valenzuela}, O., {Col{\'\i}n}, P., \& {Quinn}, T. 2009, \mnras,
  398, 1027, \dodoi{10.1111/j.1365-2966.2009.15187.x}

\bibitem[{{Lansbury} {et~al.}(2014){Lansbury}, {Lucey}, \&
  {Smith}}]{lansbury_et_al_2014}
{Lansbury}, G.~B., {Lucey}, J.~R., \& {Smith}, R.~J. 2014, \mnras, 439, 1749,
  \dodoi{10.1093/mnras/stu049}

\bibitem[{{Le Conte} {et~al.}(2024){Le Conte}, {Gadotti}, {Ferreira},
  {Conselice}, {de S{\'a}-Freitas}, {Kim}, {Neumann}, {Fragkoudi},
  {Athanassoula}, \& {Adams}}]{le_conte_et_al_2024}
{Le Conte}, Z.~A., {Gadotti}, D.~A., {Ferreira}, L., {et~al.} 2024, \mnras,
  530, 1984, \dodoi{10.1093/mnras/stae921}

\bibitem[{{Lee} {et~al.}(2012){Lee}, {Park}, {Lee}, \& {Choi}}]{lee_et_al_2012}
{Lee}, G.-H., {Park}, C., {Lee}, M.~G., \& {Choi}, Y.-Y. 2012, \apj, 745, 125,
  \dodoi{10.1088/0004-637X/745/2/125}

\bibitem[{{Lee} {et~al.}(2019){Lee}, {Ann}, \& {Park}}]{lee+ann+park_2019}
{Lee}, Y.~H., {Ann}, H.~B., \& {Park}, M.-G. 2019, \apj, 872, 97,
  \dodoi{10.3847/1538-4357/ab0024}

\bibitem[{{Lemos} {et~al.}(1991){Lemos}, {Kalnajs}, \&
  {Lynden-Bell}}]{lemos_et_al_1991}
{Lemos}, J. P.~S., {Kalnajs}, A.~J., \& {Lynden-Bell}, D. 1991, \apj, 375, 484,
  \dodoi{10.1086/170210}

\bibitem[{{Li} {et~al.}(2023){Li}, {Shlosman}, {Heller}, \&
  {Pfenniger}}]{li_et_al_2023}
{Li}, X., {Shlosman}, I., {Heller}, C., \& {Pfenniger}, D. 2023, \mnras,
  \dodoi{10.1093/mnras/stad2799}

\bibitem[{{Lieb} {et~al.}(2022){Lieb}, {Collier}, \&
  {Madigan}}]{lieb_et_al_2022}
{Lieb}, E., {Collier}, A., \& {Madigan}, A.-M. 2022, \mnras, 509, 685,
  \dodoi{10.1093/mnras/stab2904}

\bibitem[{{Lin} {et~al.}(2017){Lin}, {Li}, {He}, {Xiao}, \&
  {Wang}}]{lin_et_al_2017}
{Lin}, L., {Li}, C., {He}, Y., {Xiao}, T., \& {Wang}, E. 2017, \apj, 838, 105,
  \dodoi{10.3847/1538-4357/aa657a}

\bibitem[{{{\L}okas}(2020)}]{lokas_2020gas}
{{\L}okas}, E.~L. 2020, \aap, 634, A122, \dodoi{10.1051/0004-6361/201937165}

\bibitem[{{{\L}okas}(2021)}]{lokas_2021bar}
---. 2021, \aap, 647, A143, \dodoi{10.1051/0004-6361/202040056}

\bibitem[{{Long} {et~al.}(2014){Long}, {Shlosman}, \&
  {Heller}}]{long_et_al_2014}
{Long}, S., {Shlosman}, I., \& {Heller}, C. 2014, \apjl, 783, L18,
  \dodoi{10.1088/2041-8205/783/1/L18}

\bibitem[{{L{\'o}pez} {et~al.}(2024){L{\'o}pez}, {Scannapieco}, {Cora}, \&
  {Gargiulo}}]{lopez_et_al_2024}
{L{\'o}pez}, P.~D., {Scannapieco}, C., {Cora}, S.~A., \& {Gargiulo}, I.~D.
  2024, \mnras, 529, 979, \dodoi{10.1093/mnras/stae576}

\bibitem[{{Marinova} \& {Jogee}(2007)}]{marinova+jogee_2007}
{Marinova}, I., \& {Jogee}, S. 2007, \apj, 659, 1176, \dodoi{10.1086/512355}

\bibitem[{{Marioni} {et~al.}(2022){Marioni}, {Abadi}, {Gottl{\"o}ber}, \&
  {Yepes}}]{marioni_et_al_2022}
{Marioni}, O.~F., {Abadi}, M.~G., {Gottl{\"o}ber}, S., \& {Yepes}, G. 2022,
  \mnras, 511, 2423, \dodoi{10.1093/mnras/stac105}

\bibitem[{{Mayer} \& {Wadsley}(2004)}]{mayer+wadsley_2004}
{Mayer}, L., \& {Wadsley}, J. 2004, \mnras, 347, 277,
  \dodoi{10.1111/j.1365-2966.2004.07202.x}

\bibitem[{{Melvin} {et~al.}(2014){Melvin}, {Masters}, {Lintott}, {Nichol},
  {Simmons}, {Bamford}, {Casteels}, {Cheung}, {Edmondson}, {Fortson},
  {Schawinski}, {Skibba}, {Smith}, \& {Willett}}]{melvin_et_al_2014}
{Melvin}, T., {Masters}, K., {Lintott}, C., {et~al.} 2014, \mnras, 438, 2882,
  \dodoi{10.1093/mnras/stt2397}

\bibitem[{{Michel-Dansac} \& {Wozniak}(2006)}]{michel-dansac+wozniak_2006}
{Michel-Dansac}, L., \& {Wozniak}, H. 2006, \aap, 452, 97,
  \dodoi{10.1051/0004-6361:20041038}

\bibitem[{{Michikoshi} \& {Kokubo}(2020)}]{michikoshi+kokubo_2020}
{Michikoshi}, S., \& {Kokubo}, E. 2020, \apj, 897, 65,
  \dodoi{10.3847/1538-4357/ab9369}

\bibitem[{{Miller} \& {Smith}(1979)}]{miller+smith_1979}
{Miller}, R.~H., \& {Smith}, B.~F. 1979, \apj, 227, 785, \dodoi{10.1086/156787}

\bibitem[{{Minchev} {et~al.}(2012){Minchev}, {Famaey}, {Quillen}, {Di Matteo},
  {Combes}, {Vlaji{\'c}}, {Erwin}, \& {Bland-Hawthorn}}]{minchev_et_al_2012}
{Minchev}, I., {Famaey}, B., {Quillen}, A.~C., {et~al.} 2012, \aap, 548, A126,
  \dodoi{10.1051/0004-6361/201219198}

\bibitem[{{Minchev} \& {Quillen}(2006)}]{minchev+quillen_2006}
{Minchev}, I., \& {Quillen}, A.~C. 2006, \mnras, 368, 623,
  \dodoi{10.1111/j.1365-2966.2006.10129.x}

\bibitem[{{Nair} \& {Abraham}(2010)}]{nair+abraham_2010}
{Nair}, P.~B., \& {Abraham}, R.~G. 2010, \apjl, 714, L260,
  \dodoi{10.1088/2041-8205/714/2/L260}

\bibitem[{{Newnham} {et~al.}(2020){Newnham}, {Hess}, {Masters}, {Kruk},
  {Penny}, {Lingard}, \& {Smethurst}}]{newnham_et_al_2020}
{Newnham}, L., {Hess}, K.~M., {Masters}, K.~L., {et~al.} 2020, \mnras, 492,
  4697, \dodoi{10.1093/mnras/staa064}

\bibitem[{{Nipoti} {et~al.}(2024){Nipoti}, {Caprioglio}, \&
  {Bacchini}}]{nipoti_et_al_2024}
{Nipoti}, C., {Caprioglio}, C., \& {Bacchini}, C. 2024, \aap, 689, A61,
  \dodoi{10.1051/0004-6361/202450462}

\bibitem[{{Norman} {et~al.}(1996){Norman}, {Sellwood}, \&
  {Hasan}}]{norman_et_al_1996}
{Norman}, C.~A., {Sellwood}, J.~A., \& {Hasan}, H. 1996, \apj, 462, 114,
  \dodoi{10.1086/177133}

\bibitem[{{Ostriker} \& {Peebles}(1973)}]{op_1973}
{Ostriker}, J.~P., \& {Peebles}, P.~J.~E. 1973, \apj, 186, 467,
  \dodoi{10.1086/152513}

\bibitem[{{Peschken} \& {{\L}okas}(2019)}]{peschken+lokas_2019}
{Peschken}, N., \& {{\L}okas}, E.~L. 2019, \mnras, 483, 2721,
  \dodoi{10.1093/mnras/sty3277}

\bibitem[{{Peters} \& {Kuzio de Naray}(2019)}]{peters_et_al_2019}
{Peters}, W., \& {Kuzio de Naray}, R. 2019, \mnras, 484, 850,
  \dodoi{10.1093/mnras/sty3505}

\bibitem[{{Pfenniger} \& {Friedli}(1991)}]{pfenniger+friedli_1991}
{Pfenniger}, D., \& {Friedli}, D. 1991, \aap, 252, 75

\bibitem[{{Polyachenko} {et~al.}(2016){Polyachenko}, {Berczik}, \&
  {Just}}]{polyachenko_et_al_2016}
{Polyachenko}, E.~V., {Berczik}, P., \& {Just}, A. 2016, \mnras, 462, 3727,
  \dodoi{10.1093/mnras/stw1907}

\bibitem[{{Polyachenko} \& {Just}(2015)}]{polyachenko+just_2015}
{Polyachenko}, E.~V., \& {Just}, A. 2015, \mnras, 446, 1203,
  \dodoi{10.1093/mnras/stu2171}

\bibitem[{{Rautiainen} \& {Salo}(1999)}]{rautiainen+salo_1999}
{Rautiainen}, P., \& {Salo}, H. 1999, \aap, 348, 737

\bibitem[{{Romeo} {et~al.}(2023){Romeo}, {Agertz}, \&
  {Renaud}}]{romeo_et_al_2023}
{Romeo}, A.~B., {Agertz}, O., \& {Renaud}, F. 2023, \mnras, 518, 1002,
  \dodoi{10.1093/mnras/stac3074}

\bibitem[{{Rosas-Guevara} {et~al.}(2020){Rosas-Guevara}, {Bonoli}, {Dotti},
  {Zana}, {Nelson}, {Pillepich}, {Ho}, {Izquierdo-Villalba}, {Hernquist}, \&
  {Pakmor}}]{rosas_guevara_et_al_2020}
{Rosas-Guevara}, Y., {Bonoli}, S., {Dotti}, M., {et~al.} 2020, \mnras, 491,
  2547, \dodoi{10.1093/mnras/stz3180}

\bibitem[{{Saha} \& {Elmegreen}(2018)}]{saha+elmegreen_2018}
{Saha}, K., \& {Elmegreen}, B. 2018, \apj, 858, 24,
  \dodoi{10.3847/1538-4357/aabacd}

\bibitem[{{Saha} \& {Jog}(2014)}]{saha+jog_2014}
{Saha}, K., \& {Jog}, C.~J. 2014, \mnras, 444, 352,
  \dodoi{10.1093/mnras/stu1414}

\bibitem[{{Saha} \& {Naab}(2013)}]{saha+naab_2013}
{Saha}, K., \& {Naab}, T. 2013, \mnras, 434, 1287,
  \dodoi{10.1093/mnras/stt1088}

\bibitem[{{Sanchez-Salcedo}(1999)}]{sanchez-salcedo_1999}
{Sanchez-Salcedo}, F.~J. 1999, \mnras, 303, 755,
  \dodoi{10.1046/j.1365-8711.1999.02263.x}

\bibitem[{{Sanders} {et~al.}(2024){Sanders}, {Kawata}, {Matsunaga}, {Sormani},
  {Smith}, {Minniti}, \& {Gerhard}}]{sanders_et_al_2024}
{Sanders}, J.~L., {Kawata}, D., {Matsunaga}, N., {et~al.} 2024, \mnras, 530,
  2972, \dodoi{10.1093/mnras/stae711}

\bibitem[{{Sellwood}(1980)}]{sellwood_1980}
{Sellwood}, J.~A. 1980, \aap, 89, 296

\bibitem[{{Sellwood} \& {Evans}(2001)}]{sellwood+evans_2001}
{Sellwood}, J.~A., \& {Evans}, N.~W. 2001, \apj, 546, 176,
  \dodoi{10.1086/318228}

\bibitem[{{Seo} {et~al.}(2019){Seo}, {Kim}, {Kwak}, {Hsieh}, {Han}, \&
  {Hopkins}}]{seo_et_al_2019}
{Seo}, W.-Y., {Kim}, W.-T., {Kwak}, S., {et~al.} 2019, \apj, 872, 5,
  \dodoi{10.3847/1538-4357/aafc5f}

\bibitem[{{Shen} \& {Sellwood}(2004)}]{shen+sellwood_2004}
{Shen}, J., \& {Sellwood}, J.~A. 2004, \apj, 604, 614, \dodoi{10.1086/382124}

\bibitem[{{Sheth} {et~al.}(2012){Sheth}, {Melbourne}, {Elmegreen}, {Elmegreen},
  {Athanassoula}, {Abraham}, \& {Weiner}}]{sheth_et_al_2012}
{Sheth}, K., {Melbourne}, J., {Elmegreen}, D.~M., {et~al.} 2012, \apj, 758,
  136, \dodoi{10.1088/0004-637X/758/2/136}

\bibitem[{{Sheth} {et~al.}(2008){Sheth}, {Elmegreen}, {Elmegreen}, {Capak},
  {Abraham}, {Athanassoula}, {Ellis}, {Mobasher}, {Salvato}, {Schinnerer},
  {Scoville}, {Spalsbury}, {Strubbe}, {Carollo}, {Rich}, \&
  {West}}]{sheth_et_al_2008}
{Sheth}, K., {Elmegreen}, D.~M., {Elmegreen}, B.~G., {et~al.} 2008, \apj, 675,
  1141, \dodoi{10.1086/524980}

\bibitem[{{Springel}(2005)}]{springel_2005}
{Springel}, V. 2005, \mnras, 364, 1105,
  \dodoi{10.1111/j.1365-2966.2005.09655.x}

\bibitem[{{Sylos Labini} {et~al.}(2023){Sylos Labini}, {Straccamore}, {De
  Marzo}, \& {Comer{\'o}n}}]{sylos_labini_et_al_2023}
{Sylos Labini}, F., {Straccamore}, M., {De Marzo}, G., \& {Comer{\'o}n}, S.
  2023, \mnras, 524, 1560, \dodoi{10.1093/mnras/stad1916}

\bibitem[{{Thomasson} {et~al.}(1991){Thomasson}, {Donner}, \&
  {Elmegreen}}]{thomasson_et_al_1991}
{Thomasson}, M., {Donner}, K.~J., \& {Elmegreen}, B.~G. 1991, \aap, 250, 316

\bibitem[{{Toomre}(1964)}]{toomre_1964}
{Toomre}, A. 1964, \apj, 139, 1217, \dodoi{10.1086/147861}

\bibitem[{{Toomre} \& {Toomre}(1972)}]{toomre+toomre_1972}
{Toomre}, A., \& {Toomre}, J. 1972, \apj, 178, 623, \dodoi{10.1086/151823}

\bibitem[{{Treuthardt} {et~al.}(2007){Treuthardt}, {Buta}, {Salo}, \&
  {Laurikainen}}]{treuthardt_et_al_2007}
{Treuthardt}, P., {Buta}, R., {Salo}, H., \& {Laurikainen}, E. 2007, \aj, 134,
  1195, \dodoi{10.1086/521149}

\bibitem[{{Vande Putte} {et~al.}(2009){Vande Putte}, {Cropper}, \&
  {Ferreras}}]{vande_putte_et_al_2009}
{Vande Putte}, D., {Cropper}, M., \& {Ferreras}, I. 2009, \mnras, 397, 1587,
  \dodoi{10.1111/j.1365-2966.2009.15044.x}

\bibitem[{{Vandervoort}(1982)}]{vandervoort_1982}
{Vandervoort}, P.~O. 1982, \apjl, 256, L41, \dodoi{10.1086/183792}

\bibitem[{{Vera} {et~al.}(2016){Vera}, {Alonso}, \&
  {Coldwell}}]{vera_et_al_2016}
{Vera}, M., {Alonso}, S., \& {Coldwell}, G. 2016, \aap, 595, A63,
  \dodoi{10.1051/0004-6361/201628750}

\bibitem[{{Villa-Vargas} {et~al.}(2009){Villa-Vargas}, {Shlosman}, \&
  {Heller}}]{villa-vargas_et_al_2009}
{Villa-Vargas}, J., {Shlosman}, I., \& {Heller}, C. 2009, \apj, 707, 218,
  \dodoi{10.1088/0004-637X/707/1/218}

\bibitem[{{W{\"o}lfer} {et~al.}(2023){W{\"o}lfer}, {Facchini}, {van der Marel},
  {van Dishoeck}, {Benisty}, {Bohn}, {Francis}, {Izquierdo}, \&
  {Teague}}]{wolfer_et_al_2023}
{W{\"o}lfer}, L., {Facchini}, S., {van der Marel}, N., {et~al.} 2023, \aap,
  670, A154, \dodoi{10.1051/0004-6361/202243601}

\bibitem[{{Worrakitpoonpon}(2023)}]{worrakitpoonpon_2023}
{Worrakitpoonpon}, T. 2023, \apj, 958, 128, \dodoi{10.3847/1538-4357/acf657}

\bibitem[{{Wu} {et~al.}(2021){Wu}, {Trejo}, {Espada}, \&
  {Miyamoto}}]{wu_et_al_2021}
{Wu}, Y.-T., {Trejo}, A., {Espada}, D., \& {Miyamoto}, Y. 2021, \mnras, 504,
  3111, \dodoi{10.1093/mnras/stab1087}

\bibitem[{{Zana} {et~al.}(2018){Zana}, {Dotti}, {Capelo}, {Bonoli}, {Haardt},
  {Mayer}, \& {Spinoso}}]{zana_et_al_2018}
{Zana}, T., {Dotti}, M., {Capelo}, P.~R., {et~al.} 2018, \mnras, 473, 2608,
  \dodoi{10.1093/mnras/stx2503}

\bibitem[{{Zana} {et~al.}(2022){Zana}, {Lupi}, {Bonetti}, {Dotti},
  {Rosas-Guevara}, {Izquierdo-Villalba}, {Bonoli}, {Hernquist}, \&
  {Nelson}}]{zana_et_al_2022}
{Zana}, T., {Lupi}, A., {Bonetti}, M., {et~al.} 2022, \mnras, 515, 1524,
  \dodoi{10.1093/mnras/stac1708}

\bibitem[{{Zeng} {et~al.}(2021){Zeng}, {Wang}, \& {Gao}}]{zeng_et_al_2021}
{Zeng}, G., {Wang}, L., \& {Gao}, L. 2021, \mnras, 507, 3301,
  \dodoi{10.1093/mnras/stab2294}

\bibitem[{{Zhou} {et~al.}(2020){Zhou}, {Zhu}, {Wang}, \&
  {Feng}}]{zhou_et_al_2020}
{Zhou}, Z.-B., {Zhu}, W., {Wang}, Y., \& {Feng}, L.-L. 2020, \apj, 895, 92,
  \dodoi{10.3847/1538-4357/ab8d32}

\end{thebibliography}

\end{document}